\documentclass[twocolumn,aps,prl,superscriptaddress,nofootinbib]{revtex4-2}

\usepackage{pifont}

\usepackage{color, colortbl}
\definecolor{Gray}{gray}{0.2}

\usepackage{graphicx}
\usepackage[dvipsnames,svgnames]{xcolor}
\usepackage{bm,bbm}
\usepackage{braket}
\usepackage{amsthm,amsmath,amssymb}
\usepackage{enumerate}
\usepackage{booktabs,multirow}
\usepackage{mathtools,bbding}
\usepackage[percent]{overpic}
\usepackage{microtype}
\usepackage{array,adjustbox,afterpage}
\usepackage[T1]{fontenc}
\usepackage[utf8]{inputenc}
\usepackage{tikz}
\usetikzlibrary{calc}
\usepackage[english]{babel}
\usepackage{newtxtext,newtxmath}

\usepackage{lipsum}
\usepackage{pifont}

\usepackage[linesnumbered,ruled]{algorithm2e}  

\makeatletter
\renewcommand{\fnum@figure}{FIG. \thefigure} 
\makeatother

\usepackage[bookmarks=false, colorlinks=true, linkcolor=blue, citecolor=purple, urlcolor=purple]{hyperref}
\usepackage{cleveref}
\usepackage[normalem]{ulem}

\usepackage{lipsum}
\usepackage{makecell}

\usepackage{caption}
\usepackage{subcaption}
\usepackage{ragged2e}

\Crefname{subfigures}{figure}{figures}
\Crefname{subfigures}{Figure}{Figures}

\newcommand{\xdownarrow}[1]{{\left\downarrow\vbox to #1{}\right.\kern-\nulldelimiterspace}} 

\hypersetup{
    breaklinks=true,   
    colorlinks=true,   
    pdfusetitle=true,  
}

\begin{document}

\title{Thermodynamic limits of the Mpemba effect: \\A unified resource theory analysis of correlation-enabled mechanisms}
\author{Doruk Can Aly\"{u}r\"{u}k}
\email{dalyuruk@stu.khas.edu.tr}
\affiliation{Faculty of Engineering and Natural Sciences, Kadir Has University, 34083, Fatih, Istanbul, T\"{u}rkiye}

\author{Mahir H. Ye\c{s}iller}
\affiliation{T\"UB\.ITAK Informatics and Information Security Research Center (B\.ILGEM), 41470, Gebze, Kocaeli, T\"{u}rkiye}

\author{Vlatko Vedral}
\affiliation{Department of Physics, University of Oxford, Parks
Road, Oxford, OX1 3PU, UK}

\author{Onur Pusuluk}
\email{onur.pusuluk@gmail.com}
\affiliation{Faculty of Engineering and Natural Sciences, Kadir Has University, 34083, Fatih, Istanbul, T\"{u}rkiye}

\begin{abstract}

The Mpemba effect, in which a hotter system cools faster than a colder one, remains one of the most intriguing anomalies in thermodynamics. Here, we investigate its microscopic origin within the framework of quantum resource theories and introduce correlations as a new enabling mechanism: classical correlations can support the effect, whereas quantum correlations become relevant only under specific energy-degeneracy conditions. Importantly, correlations are necessary but not sufficient. Whether they induce the effect depends on their distribution across subsystems and on system parameters. Other resources, such as non-Markovian memory effects and Hilbert space dimensionality, primarily modulate the temperature window in which the effect can occur. Finally, by analyzing both didactic multi-qubit instances and a phenomenological single-molecule model of water, we demonstrate that the insufficiency of correlations helps explain the sporadic and sometimes contradictory observations of the Mpemba effect in experiments.

\end{abstract}

\maketitle

\textit{Introduction.}--- Common intuition suggests that hotter systems cool more slowly than cooler ones under identical conditions. However, this assumption does not always hold. Remarkably, under certain circumstances, hot water has been observed to freeze faster than cold water -- a thermodynamic anomaly known as the Mpemba effect (ME). This phenomenon was first formalized by Mpemba and Osborne~\cite{mpemba1969cool, kell1969freezing}, though its origins trace back centuries, with references appearing in the works of Aristotle~\cite{aristotle_meteorology}, Descartes~\cite{descartes_discourse}, and Bacon~\cite{bacon1878novum}.

In recent years, anomalous cooling has been documented well beyond water, in systems such as quenched polymers~\cite{hu2018conformation}, clathrate hydrates~\cite{ahn2016experimental}, and colloidal suspensions~\cite{2020_Nature, kumar2022anomalous}. Moreover, the anomaly is not limited to cooling: related relaxation speedups have been reported in magnetic alloys~\cite{chaddah2010overtaking}, spin models~\cite{baity2019mpemba,yang2020non,vadakkayil2021should,teza2023relaxation}, systems approaching equilibrium without phase transitions~\cite{lu2017nonequilibrium,klich2019mpemba,gal2020precooling,walker2021anomalous,busiello2021inducing}, and driven molecular gases relaxing to nonequilibrium steady states~\cite{lasanta2017hotter,biswas2020mpemba,takada2021mpemba,mompo2021memory,megias2022thermal,biswas2022mpemba}. Importantly, analogous phenomena have been identified in quantum domains~\cite{2025_arXiv_review, 2025_arXiv_review_q}, from integrable~\cite{rylands2024microscopic,rylands2024dynamical,chalas2024multiple,murciano2024entanglement} and chaotic~\cite{liu2024symmetry,wang2024mpembachaos} models to quantum dots~\cite{chatterjee2023quantum,graf2024role}, and recently demonstrated in trapped-ion experiments~\cite{joshi2024observing, aharony2024inverse, zhang2025observation}. This breadth motivates a unifying operational account.

Despite its ubiquity, the mechanisms driving the ME remain elusive. Proposed explanations are diverse~\cite{auerbach1995supercooling,zhang2014hydrogen,sun2015temperature,tyrovolas2017jmp,2017_JCTC_HBond,jin2015mechanisms,vynnycky2015can,lu2017nonequilibrium,bechhoefer2021fresh}, but no comprehensive framework has emerged. In quantum settings, attempts to explain the ME commonly rely on master-equation models~\cite{nava2019lindblad,carollo2021exponentially,manikandan2021equidistant,bao2022accelerating,kochsiek2022accelerating,ivander2023hyperacceleration, chatterjee2024multiple, wang2024mpemba, caceffo2024entangled, strachan2024non, nava2024mpemba, moroder2024thermodynamics, longhi2024bosonic, longhi2024mpemba, medina2024anomalous, qian2024intrinsic}. Here, we instead adopt the resource-theoretic view~\cite{coecke2016mathematical,chitambar2019quantum} of athermality~\cite{gour2015resource,goold2016role,lostaglio2019introductory,torun2023compendious}, which recognize not only temperature but also non-Markovian memory effects~\cite{yunger2020fundamental,spaventa2022capacity}, initial correlations~\cite{burkhard2024boosting}, and system dimensionality~\cite{tiwary2024quantum} as thermal resources governing relaxation. While recent resource-theoretic studies of the ME~\cite{2024_ThermoMajMPemba,2025_arXiv_TCD} have made progress, they did not address the role of such additional thermal resources.

\begin{figure} [b]
    \centering
    \includegraphics[width=.95\columnwidth]{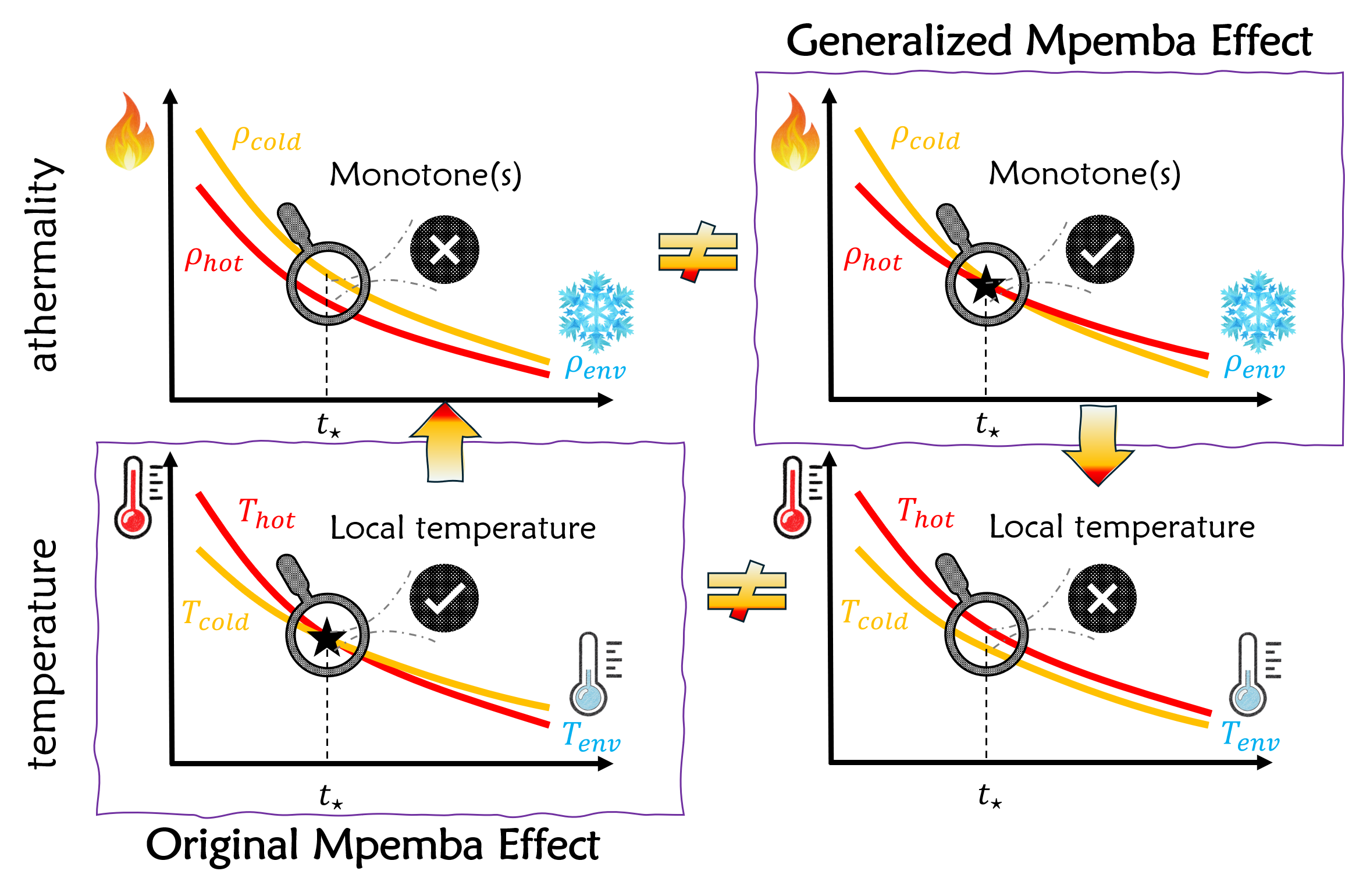}
    \caption{\justifying  Correlation-enabled manifestations of the thermal ME. (i) Faster cooling of hotter systems without monotone crossings, and (ii) faster relaxation of states farther from equilibrium without temperature crossings. Other resources, such as non-Markovianity and dimensionality, act primarily as secondary amplifiers, while the effect itself depends sensitively on the distribution of correlations and on spectral structure.
    }
    \label{fig:defs}
\end{figure}

This broader framework of athermality allows us to disentangle two notions often conflated: (i) the faster cooling of hotter systems (\textit{original} ME), and (ii) the faster relaxation of states farther from equilibrium (\textit{generalized} ME). We show that correlations can render a colder system farther from equilibrium than a hotter one, thereby producing two inequivalent manifestations: correlation-enabled original and generalized MEs (Fig.~\ref{fig:defs}). In both cases, correlations are necessary but not sufficient, while non-Markovianity and dimensionality act primarily as secondary amplifiers. The effectiveness of correlations depends sensitively on their distribution and on the structure of the energy eigenspectrum, which may explain why anomalous cooling is observed only sporadically across experiments. Our results thus extend the thermodynamic role of correlations beyond anomalous heat flows~\cite{2008_PRE_Partovi,2009_PRL_qArrowOfTime,2010_PRE_JenningsAndRudolph,daug2016multiatom,manatuly2019collectively,latune2019heat,micadei2019reversing,2020_PRA_HeatFlowReversalIonTrap,pusuluk2021quantum,lipka2024fundamental,2025_PRL_Alex}, establishing them as a unifying mechanism that also governs anomalous cooling and relaxation across classical and quantum domains.

\textit{Definitions.}---
The definition of thermal ME has two components: the assignment of a temperature to nonequilibrium states and the description of their thermal relaxation.

\textit{Definition 1.} (Temperature assignment.)---
A system with Hamiltonian \(\hat H\) is Gibbsian at temperature \(T_x\) if its state is
\(\hat\rho_{\mathrm{th}}(T_x,\hat H)=e^{-\beta_x \hat H}/Z_x(\hat H)\), with \(\beta_x=1/k_BT_x\) and \(Z_x(\hat H)=\mathrm{tr}[e^{-\beta_x \hat H}]\).
Each Gibbs state is in equilibrium with respect to its own temperature but out of equilibrium relative to a bath at \(T_b\neq T_x\). More generally, for \(\hat H=\sum_j\hat H_j\), a joint state
\(\hat\rho_{\mathrm{corr}}(T_x)\) is \emph{locally thermal} at \(T_x\) if every marginal is Gibbsian, \(\hat\rho_k=\mathrm{tr}_{\{1,\ldots,n\}\setminus\{k\}}[\hat\rho_{\mathrm{corr}}(T_x)]
=\hat\rho_{\mathrm{th}}(T_x,\hat H_k)\),
even though the global state is not a product. These \emph{locally thermal correlated states} are, by definition, correlated as
\(I(1:\cdots:n)=S(\hat\rho_{\mathrm{corr}}\|\hat\rho_1\otimes\cdots\otimes\hat\rho_n)>0\), and such correlations can affect the temperature read by a collisional thermometer through reciprocal heat–information relations~\cite{pusuluk2021quantum}.

\textit{Definition 2.} (Hierarchy of relaxations)---
Thermal relaxation forms a strict hierarchy:
(i) \emph{Gibbs-preserving maps} (GPMs) leave the Gibbs state at the bath temperature \(T_b\) invariant,  
(ii) \emph{thermal operations} (TOs) are those implementable by energy-conserving unitaries governing system-bath interactions~\cite{FundLimNano2013},  
(iii) \emph{elementary TOs} (ETOs) act only on two levels at a time~\cite{2018_Quantum_eTOs}, and 
(iv) \emph{Markovian TOs} (MTOs) correspond to Lindblad semigroups~\cite{ContThermo2022}. Thus, we have the inclusion chain
\(
\text{MTOs} \;\subset\; \text{ETOs} \;\subset\; \text{TOs} \;\subset\; \text{GPMs}
\).

The appropriate description for a given experimental realization of anomalous relaxation depends on the constraints of the setup, which may restrict the accessible operations to a particular level of the hierarchy. In this Letter, we focus on TOs and their Markovian subclass, as the physical realizability of general GPMs is subtle~\cite{2015_NJP_GibbsPrervingMapsVsTOs, 2025_PRL_GibbsPvsTO}. For block-diagonal states in the energy eigenbasis, the generalization of the second law to single systems under TOs is fully characterized by thermo-majorization~\cite{FundLimNano2013, MeadRuch1976, AMead1977}, while generalized R\'{e}nyi free energies \(F_\alpha\) provide necessary athermality monotones~\cite{SecLaws2015}. For MTOs, the stronger framework of \emph{continuous} thermo-majorization applies~\cite{ContThermo2022}. For completeness, detailed definitions of these tools are provided in the Supplemental Material (SM).

\textit{Theorem 1.} (No-go for Gibbs states)---
Fix a bath at \(T_b\) and consider a system with Hamiltonian \(\hat H\). For Gibbs states \(\hat\rho_{\mathrm{th}}(T_h,\hat H)\) and \(\hat\rho_{\mathrm{th}}(T_c,\hat H)\) with \(T_h>T_c>T_b\), (continuous) thermo-majorization guarantees that
\(\hat\rho_{\mathrm{th}}(T_c,\hat H)\) is always closer to equilibrium than
\(\hat\rho_{\mathrm{th}}(T_h,\hat H)\).
Thus, colder-but-farther configurations resulting in MEs are impossible within the product Gibbs family under (Markovian) TOs.

\textit{Proof sketch.}---
Immediate from thermo-majorization (or its continuous version), since Gibbs states define a totally ordered curve relative to \(\hat\rho_{\mathrm{th}}(T_b,\hat H)\).
Full details are given in the SM.

\textit{Remark.}---
Colder-but-farther scenarios occur only when the cold sample is in a correlated nonequilibrium state \(\hat{\rho}_{\mathrm{cold}}=\hat{\rho}_{\text{corr}}(T_c)\) with locally thermal marginals at \(T_c\), since appropriate correlations can render possible the ordering reversal \(F_\alpha[\hat{\rho}_{\rm cold}] > F_\alpha[\hat{\rho}_{\rm hot}]\) for all \(\alpha\) that underlies both original and generalized MEs. The effect is most pronounced if the state of the hot sample is a Gibbsian \(\hat{\rho}_{\mathrm{hot}}=\hat{\rho}_{\text{th}}(T_h)\) with \(T_h>T_c\).

Fig.~\ref{fig:defs} illustrates these correlation-enabled scenarios in terms of temperature and monotone trajectories. Explicit master-equation examples are given in the SM (Sec.~\ref{sec:SM:CorrME}), showing that correlations can indeed generate all these types of trajectories. These equations produce GPMs that are not necessarily genuine TOs; hence, their implementability requires a microscopic derivation. We therefore adopt a resource-theoretic perspective, using thermo-majorization to capture the role of correlations and to characterize the thermodynamic limits of that role independently of specific dynamical models.

\textit{Definition 3.} (Mpemba window)---
Let \(\mathcal{C}(T_c)\) be the set of block-diagonal joint states with marginals
Gibbs at \(T_c\). Define
\[
T_h^{\max}(T_c) = \sup\{\,T_h \;|\; \exists\,\hat{\rho}_{\mathrm{corr}}\in\mathcal{C}(T_c):
\hat{\rho}_{\mathrm{corr}} \succ_{T_b} \hat{\rho}_{\mathrm{th}}(T_h,\hat H)\,\},
\]
where \(\succ_{T_b}\) is the thermo-majorization preorder relative to the Gibbs state \(\hat{\rho}_{\mathrm{th}}(T_b,\hat H)\). The quantity \(T_h^{\max}(T_c)\) therefore characterizes the maximal hot temperature against which a colder correlated state can still be farther from equilibrium at \(T_b\). The appropriate comparison between \(\hat{\rho}_{\mathrm{corr}}(T_c)\) and
\(\hat{\rho}_{\mathrm{th}}(T_h,\hat H)\) depends on the operational restrictions of the relaxation process, which in practice are set by the experimental constraints of the setup. For example, with catalysts the ordering must be defined through the family
\(\{F_\alpha\}\), whereas for memoryless (Markovian) processes the relevant relation
is continuous thermo-majorization.

\textit{Proposition 1.}---
For any TO at bath temperature \(T_b\), if a correlation-enabled ME occurs between a locally thermal correlated state at \(T_c\) and a Gibbs state at \(T_h\), then necessarily \(T_h \leq T_h^{\max}(T_c)\).

\textit{Interpretation.}---
The bound \(T_h^{\max}(T_c)\) provides a dynamics-independent constraint on the
observable hot temperature, defining a window \([T_c,\,T_h^{\max}(T_c)]\) for correlation-enabled mechanisms. If \(T_h > T_h^{\max}(T_c)\), both original and generalized effects are forbidden. For \(T_h \leq T_h^{\max}(T_c)\), different outcomes are possible: original, generalized, or neither effect may occur. Thus, the window identifies the maximal temperature range in which any
correlation-enabled manifestation of the ME can take place, without committing to which type is realized.

\textit{Main result.}---
Theorem~1 and Proposition~1 together show that correlations are \emph{necessary but not sufficient} for colder-but-farther configurations resulting in MEs. Their absence rules out these new manifestations entirely, while their presence alone does not guarantee them:
realization depends sensitively on microscopic parameters and on how correlations are distributed. This sensitivity, observed both in didactic multi-qubit instances and in phenomenological single-molecule models of water, may explain why experimental observations of the ME are sporadic and sometimes contradictory. Other resources, such as non-Markovian memory effects and system dimensionality, play only a secondary role by modulating the maximal temperature window in which the effect may occur.

\textbf{Result 1.}---
Classical correlations shared between subsystems in local thermal equilibrium can give rise to the ME.

For illustration, we consider identical qubit systems; the generality of the result does not depend on system dimension (see also the water example below). Each qubit has Hamiltonian \(\hat{H}_q = E_g |g\rangle \langle g| + E_e |e\rangle \langle e|\).  In the hot sample, all \(n\) qubits are in a product state, \(\hat{\rho}_{\rm hot} = \hat{\rho}_P(n, T_h) = \hat{\rho}_{\rm th}(T_h, \hat{H}_q)^{\otimes n}\). In the cold sample, \(m\) out of \(n\) qubits are prepared in a classically correlated local thermal state,  \(\hat{\rho}_{C}(m, T_c) = P_g(T_c)\, (|g\rangle\langle g|)^{\otimes m} + P_e(T_c)\,(|e\rangle\langle e|)^{\otimes m}\), with \(P_{g/e}(T_c) = e^{-\beta_c E_{g/e}}/Z_c(\hat{H}_q)\). The remaining \(n-m\) qubits are in the product state at the same temperature, giving \(\hat{\rho}_{\rm cold} = \hat{\rho}_{C}(m, T_c) \otimes \hat{\rho}_P(n-m, T_c)\).
 \begin{figure} [t]
    \centering
    \includegraphics[width=\columnwidth]{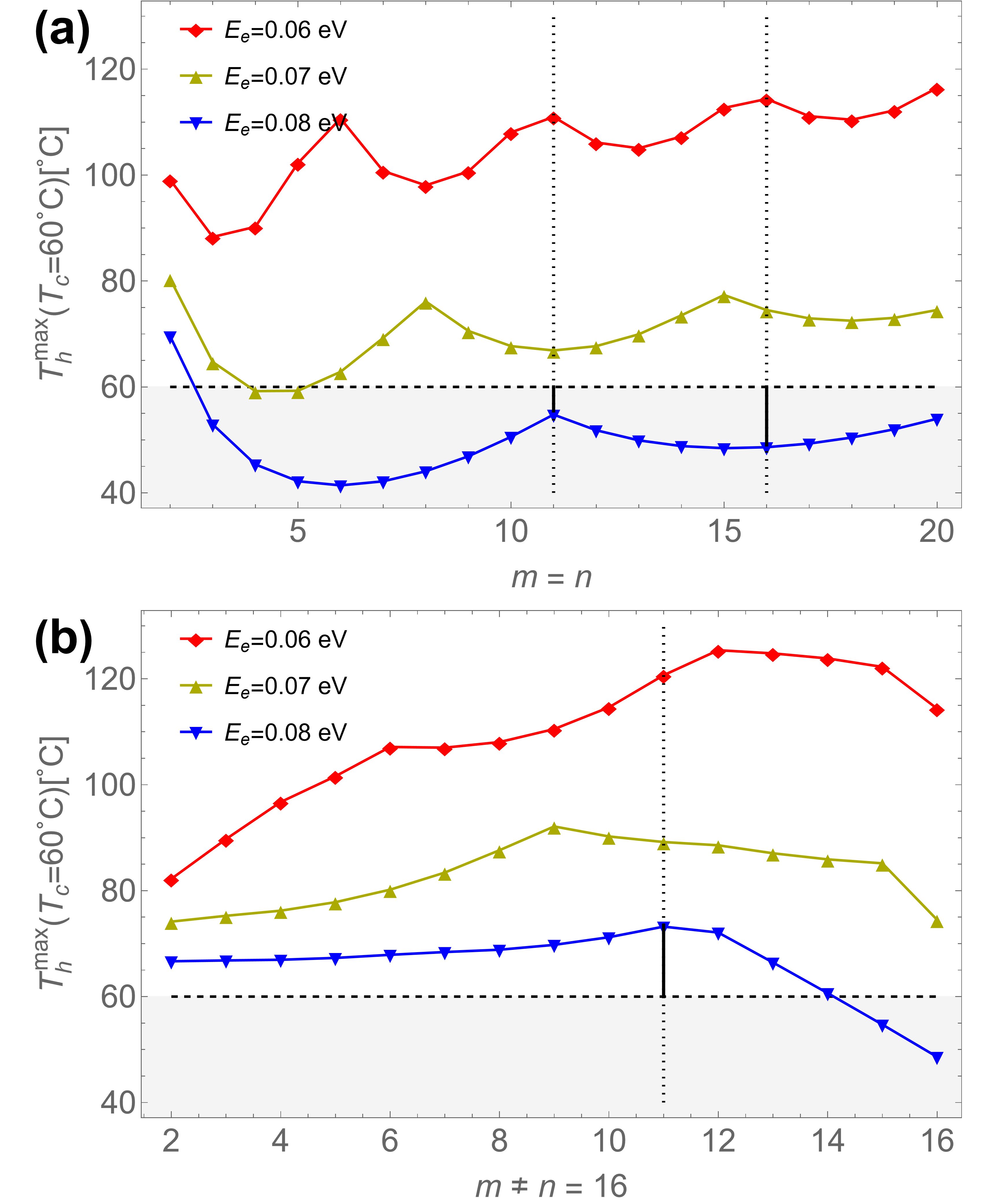}
    \caption{\justifying Sensitivity of the Mpemba windows to microscopic parameters and correlation patterns in multi-qubit systems at \(T_c=60^\circ\)C, obtained by comparing \(\hat{\rho}_{\rm cold} = \hat{\rho}_{C}(m,T_c)\otimes \hat{\rho}_P(n-m,T_c)\) with \(\hat{\rho}_{\rm hot} = \hat{\rho}_P(n,T_h)\), where \(n\) and \(m\) are the numbers of total and correlated qubits.  Results are shown for different energy gaps: \(E_e=0.06\) eV (red \(\blacklozenge\)), \(0.07\) eV (green \(\blacktriangle\)), and \(0.08\) eV (blue \(\blacktriangledown\)).  The horizontal dashed line marks \(T_c\). The shaded region below this line highlights parameter regimes where the correlation-enabled MEs is absent, while data points above indicate values of \(T_h\) for which the effect is realized. In the upper panel, all qubits share correlations (\(n=m\)) whereas in the lower panel only a subset of \(m\leq n\) qubits are correlated.}
    \label{fig:multiC}
\end{figure}

As a first step, consider the simple two-qubit case. Restricting the set \(\mathcal{C}(T_c)\) in Definition 3 to \(\hat{\rho}_{\rm corr}=\hat{\rho}_{C}(2,T_c)\), the maximal hot temperature \(T_h^{\max}(T_c)\) follows from the conditions \(e^{-2\beta_h E_e}/Z_P < e^{-\beta_c E_e}/Z_C\) and \((e^{-2 \beta_h E_e}+ 2 e^{-\beta_h(E_g+E_e)})/Z_P
= (e^{-\beta_c E_e}+ 2 e^{-\beta_b(E_e-E_g)-\beta_c E_g})/Z_C\), where \(Z_C = Z_c(\hat{H}_q)\) and \(Z_P = Z_h(\hat{H}_q \otimes \mathbb{I} + \mathbb{I} \otimes \hat{H}_q)\) (see the SM for an illustrative derivation using thermo-majorization curves). Solving these equations yields the Mpemba window \([T_c,\,T_h^{\max}(T_c)]\). For example, with \(E_g=0\) eV and \(E_e=0.05\) eV, the correlated state at \(60^\circ\)C admits a window extending up to \(T_h^{\max}=136.7^\circ\)C during relaxation toward \(0^\circ\)C. Increasing the gap shrinks this window: for \(E_e=0.06,\,0.07,\) and \(0.08\) eV we obtain
\(T_h^{\max}\approx 99.12^\circ\)C, \(80.40^\circ\)C, and \(69.56^\circ\)C, respectively.

The multi-qubit case reveals a richer structure. Fig.~\ref{fig:multiC} shows the maximal \(T_h\) achievable when comparing 
\(\hat{\rho}_{\rm cold} = \hat{\rho}_{C}(m,T_c)\otimes \hat{\rho}_P(n-m,T_c)\) with 
\(\hat{\rho}_{\rm hot} = \hat{\rho}_P(n,T_h)\). For \(E_e=0.06\) eV and \(0.07\) eV the effect persists for many values of \(n\) and \(m\), while for \(E_e=0.08\) eV it disappears whenever \(n=m\ge 2\) (Fig.~\ref{fig:multiC}~a). Remarkably, the effect reemerges when correlations are confined to a subset of qubits.  
As shown by the vertical lines in Fig.~\ref{fig:multiC}~a, it is absent when all \(n=11\) or \(n=16\) qubits are correlated, yet reappears when only \(11\) out of \(16\) share correlations (Fig.~\ref{fig:multiC}~b).

The correlations in the initial cold sample could be distributed in different ways. For \(T_c = 60^\circ\)C, pairwise correlations in \(\hat{\rho}_{\rm cold} = \hat{\rho}_{C}(2, T_c)^{\otimes n/2}\) make the Mpemba window widen almost exponentially with system size \(n\) across all examined energy gaps (see Fig.~\ref{fig:SM:DimerizedMultiC} in the SM). In contrast, a single correlated pair, \(\hat{\rho}_{\rm cold} = \hat{\rho}_{C}(2, T_c) \otimes \hat{\rho}_{P}(n-2, T_c)\), exhibits a nonmonotonic dependence on dimensionality (Fig.~\ref{fig:SM:SingleMultiC}). These results show that classical correlations can enable colder-but-farther configurations whose thermodynamic advantage depends not only on their total strength but also on how correlations are distributed across subsystems.


\textbf{Result 2.} Quantum correlations shared between subsystems in local thermal equilibrium may result in the ME only under specific spectral conditions arising from energy degeneracies.

Any density matrix can be decomposed in the energy basis into coherence modes~\cite{2014_PRA_ModesOfCoh,2015_PRX_ModesOfCoh},
\begin{equation}
       \hat{\rho} = \sum_{\omega \, \in \, \Omega} \hat{\rho}^{(\omega)} \,, \quad \hat{\rho}^{(\omega)}  \equiv \sum_{E_n-E_m=\omega} \rho_{nm} |n \rangle \langle m| ,
\end{equation}
where \(\Omega\) is the Bohr spectrum. Each mode then evolves independently under TOs, \(
\mathcal T( \hat{\rho}^{(\omega)} ) =  \hat{\sigma}^{(\omega)} \equiv \mathcal T(\hat{\rho} )^{(\omega)}\). Energy populations belong exclusively to the zero mode (\(\omega = 0\)), which also contains coherences between degenerate energy levels. Thus, only correlations in the zero mode co-evolve with populations and affect their relaxation dynamics, while correlations in nonzero modes neither couple to heat exchange nor influence populations, but simply decay away. Both scenarios can be illustrated using qubit systems, as in Result 1.

As a representative case, consider a cold sample prepared in \(\hat{\rho}_{\rm cold} = \hat{\rho}_{E}(m, T_c) \otimes \hat{\rho}_P(n-m, T_c)\), where \(m\) out of \(n\) qubits share multipartite entanglement in \(\hat{\rho}_E(m, T_c) = \hat{\rho}_C(m, T_c) + \mu \, (|g\rangle \langle e|)^{\otimes m}+ \mu^* (|e \rangle \langle g|)^{\otimes m}\) with \(|\mu| \leq \sqrt{P_g(T_c) \, P_e(T_c)}\). The off-diagonal elements of \(\hat{\rho}_{\rm cold}\) belong to the nonzero coherence modes \(\omega = m (E_e - E_g)\). Under TOs, coherences in these modes evolve independently of the energy populations. Since \(\hat{\rho}_E(m, T_c)\) and \(\hat{\rho}_C(m, T_c)\) contain the same classical correlations, the ME persists under the same conditions. However, catalysts can, in principle, activate correlations in non-degenerate subspaces to mediate effective heat exchange with the bath~\cite{2025_PRL_Alex}, and thus extend the temperature window for \(\hat{\rho}_{E}(m, T_c) \otimes \hat{\rho}_P(n-m, T_c)\) relative to its classically correlated counterpart.

Different correlation patterns lead to further variations. In particular, if entanglement is shared pairwise, \(\hat{\rho}_{\rm cold}=\hat{\rho}_{E}(2, T_c)^{\otimes m/2} \otimes \hat{\rho}_P(n-m, T_c)\), then for \(m \geq 4\) the tensor-product structure can generate zero-mode coherence contributions, allowing the Mpemba window to exceed that of the corresponding classically correlated pairwise configurations. This enhancement grows rapidly as more entangled pairs are added, and in the fully dimerized case \(m=n\) it is absent at \(n=2\) but becomes positive for \(n\geq 4\), increasing overall with system size (see Figs.~\ref{fig:SM:EntPairwisePartial}~and~\ref{fig:SM:EntVsClassicalDimerized} in the SM).

A complementary scenario considers a cold sample prepared in \(\hat{\rho}_{\rm cold} = \hat{\rho}_{D}(m, T_c) \otimes \hat{\rho}_P(n-m, T_c)\), where \(m\) out of \(n\) qubits form a multipartite discordant state, \(\hat{\rho}_D(m, T_c) = \hat{\rho}_P(m,T_c) + \lambda \, (|g e\rangle \langle e g|)^{\otimes m/2} + \lambda^* (|e g\rangle \langle g e|)^{\otimes m/2}\), with \(|\lambda|\leq (P_g(T_c) \, P_e(T_c))^{m/2}\). All coherences in this state lie in the zero mode, which can dynamically interconvert with energy-level populations under TOs.

For the two-qubit case, the ordering reversal occurs between the cold and hot samples when \((\beta_b - \beta_c) \, (E_e-E_g) \leq \ln{2}\) at maximal \(\lambda\). Setting the single-qubit energy levels to \(E_g = 0\) eV and \(E_e = 0.05\) eV, the cold sample at \(60^\circ\)C becomes farther from equilibrium than the hot sample with a temperature as high as \(99.62^\circ\)C during relaxation toward \(0^\circ\)C. More generally, for the discordant families considered here, increasing either the size of the discordant sector or the total system size narrows the Mpemba window across all energy gaps examined, driving the maximal hot temperature back toward the trivial baseline \(T_h^{\max}\approx T_c\) (see Figs.~\ref{fig:SM:DiscordAll}~and~\ref{fig:SM:DiscordEmbedded} in the SM).

Interestingly, this suppression is not generic to all discord distributions: when the maximum discord is distributed pairwise among all qubits, e.g., \(\hat{\rho}_{\rm cold} = \hat{\rho}_{D}(2, T_c)^{\otimes n/2}\), the Mpemba window remains exactly equal to that of the two-qubit discordant state \(\hat{\rho}_{D}(2, T_c)\) for all even \(n\) (see the SM). All these examples demonstrate that quantum correlations can also create colder-but-farther configurations leading to the ME. Unlike classical correlations, their impact is less constrained by the fine-grained spectral structure and more sensitive to how correlations are distributed and structured, with zero-mode participation playing the central role.

\textbf{Result 3.} Building on Results 1 and 2, we now examine the role of Hilbert-space dimensionality. Previous works have established it as a thermodynamic resource in its own right~\cite{2016_Dim,tiwary2024quantum}. In certain systems, including the one analyzed in Ref.~\cite{burkhard2024boosting}, dimensionality can even outweigh correlations, although this aspect was not addressed in that study. Our findings above show that, for correlation-enabled MEs, dimensionality is only secondary and highly non-universal: increasing system size usually narrows the Mpemba window, though in rare cases it can transiently expand it. Thus, dimensionality does not provide an independent mechanism for the effect, but merely modulates how correlations affect relaxation.

\textbf{Result 4.} Non-Markovian memory effects broaden the temperature window within which the ME is sustained by correlations. 

As noted in Definition 3, while TOs are characterized by thermo-majorization, their Markovian subclass (MTOs) requires continuous thermo-majorization~\cite{ContThermo2022}. For the classically correlated state \(\hat{\rho}_{\rm cold}=\hat{\rho}_C(2,60^\circ\text{C})\), the maximal hot temperature drops from \(136.7^\circ\text{C}\) to \(102.8^\circ\text{C}\) when the thermal relaxation is restricted to MTOs. For the discordant state \(\hat{\rho}_{\rm cold}=\hat{\rho}_D(2,60^\circ\text{C})\), it decreases from \(99.62^\circ\text{C}\) to \(95.5^\circ\text{C}\) under the same restriction. These results, illustrated with continuous thermo-majorization curves Figs.~\ref{fig:SM:2qCcont}~and~\ref{fig:conttmaj_pd} in the SM, show that non-Markovianity can widen the range over which both classical and quantum correlations support the ME.


\textit{Mpemba effect in water.}---
The ME in water has often been attributed to hydrogen-bond (HB) dynamics, with explanations ranging from bond-memory effects and entropy differences to altered strong/weak HB ratios~\cite{zhang2014hydrogen,sun2015temperature,tyrovolas2017jmp,2017_JCTC_HBond}, though such mechanisms remain debated~\cite{burridge2016srep}. On the other hand, water molecules can transiently establish quantum correlations (discord and entanglement) via proton delocalization along HBs~\cite{Pusuluk_2018_PRSA,Pusuluk_2019_PRSA}, which rapid molecular motion typically erases, leaving only classical correlations. Motivated by these observations and controversies, we develop a phenomenological single-molecule model to test whether classical correlations between HB numbers can account for the ME.

A molecule can form up to four HBs. We assume the energy of the \(j\)th bond at temperature \(T_x\) is 
\(\epsilon(j,T_x) = \epsilon_0(1-\alpha T_x) + \gamma j\), 
where \(\epsilon_0\) sets the baseline HB strength, \(\alpha\) describes thermal weakening, and \(\gamma\) encodes cooperative effects. 
The energy of a configuration with \(N\) bonds is then \(E_N(T_x) = N\epsilon_0(1-\alpha T_x) + \tfrac{\gamma}{2}N(N+1)\). To capture structural heterogeneity, we refine the extremes of the bonding spectrum. At the fully bonded end, we distinguish between a distorted tetrahedral state (\(4_{\rm L}\)) and an ideal ice-like geometry (\(4_{\rm I}\)), with the latter stabilized as  
\(E_{4_{\rm I}}(T_x) = E_{4_{\rm L}}(T_x) - \Delta q/(1+e^{(T_{L\leftrightarrow I}-T_x)})\). At the unbonded end, we separate liquid-like isolated molecules (\(0_{\rm L}\)) from vapor-like ones (\(0_{\rm V}\)), weighted entropically by
\(w=1+\Delta_{\rm vap}/(1+e^{(T-T_{L\leftrightarrow V})})\). This construction yields a 7-level model, 
\(\hat{H}_{\rm H_2O} = \sum_{j} E_j |E_j\rangle\langle E_j|\) 
with \(j \in \{0_{\rm V},0_{\rm L},1_{\rm L},2_{\rm L},3_{\rm L},4_{\rm L},4_{\rm I}\}\).

As shown in didactic multi-qubit instances, small spectral shifts crucially affect relaxation. Accordingly, we fit the model parameters to TIP4P data~\cite{jorgensen1985watersim} (\(\epsilon_0=-0.66931\,\mathrm{eV}\), 
\(\alpha=7.097\times10^{-4}\,\mathrm{K}^{-1}\), 
\(\gamma=0.13436\,\mathrm{eV}\), 
\(\Delta q=0.09519\,\mathrm{eV}\), 
and \(\Delta_{\rm vap}=6.03\times10^9\)), ensuring that our single-molecule model reproduces key temperature-dependent structural and thermodynamic observables, with smooth \(L\leftrightarrow I\) and \(L\leftrightarrow V\) crossovers instead of sharp bulk transitions (see SM).

With this parametrized model in hand, we next test whether classical correlations can account for the ME. As in Result~1, we compare correlated and uncorrelated preparations of water molecules at the same initial temperature. We denote by \(\hat{\rho}_{P}(({\rm H_2O})_n,T_x) \equiv  \hat{\rho}_{\rm th}(T_x, \hat{H}_{\rm H_2O})^{\otimes n}\) the product of \(n\) single-molecule Gibbs states of the 7-level Hamiltonian, and by \(\hat{\rho}_C(({\rm H_2O})_n,T_x) \equiv \sum_j P_j(T_x) (|E_j\rangle \langle E_j|)^{\otimes n}\) with \(P_j(T_x) = e^{-\beta_x E_j}/Z_x({\rm H_2O})\) the \(n\)-partite classically correlated mixture of local thermal states. The correlated cold sample is then
\(\hat{\rho}_{\rm cold} = \hat{\rho}_C(({\rm H_2O})_m,T_c)\otimes \hat{\rho}_P(({\rm H_2O})_{n-m},T_c)\), while the hot sample is simply the uncorrelated product \(\hat{\rho}_{\rm hot} = \hat{\rho}_P(({\rm H_2O})_n,T_h)\). Comparing \(\hat{\rho}_{\rm cold}\) and \(\hat{\rho}_{\rm hot}\) under TOs identifies the maximal hot temperature \(T_h^{\max}(T_c)\) at which the correlated cold state remains farther from equilibrium than its hotter counterpart.

Table~\ref{tab:watercmn} reports the width of the Mpemba window, $\Delta T_{\rm M}=T_h^{\max}-T_c$, across different values of $(n,m)$ and $T_c$. Several systematic trends emerge. First, the window is largest at low $T_c$ and shrinks as $T_c$ increases, disappearing above $\sim 80^\circ$C in most cases. Second, increasing the correlated fraction $m$ usually enlarges the window, though the dependence can be non-monotonic and intertwined with system size $n$. Overall, the 7-level water model confirms the phenomenological picture established in Results~1 and~2: initial correlations can enable colder-but-farther configurations, but the strength of the effect depends sensitively on temperature, system size, and the distribution of correlations. This pronounced variability provides a natural rationale for the sporadic and sometimes contradictory reports of the ME in water experiments.

\begin{table}[t]
\caption{\justifying
Width of the Mpemba window, \(\Delta T_{\rm M}=T_h^{\max}-T_c\) (°C), in the 7-level water model under TO with bath temperature fixed at \(T_b = 0.8^\circ\)C, the lowest temperature at which the liquid regime persists in the model. Rows show the correlated fraction \((n,m)\), with \(m\) out of \(n\) molecules sharing classical correlations. Columns indicate the initial cold temperature \(T_c\). A cross (×) marks cases with no ME, i.e., \(T_h^{\max}\!\le T_c\).
}
\label{tab:watercmn}
\begin{ruledtabular}
\begin{tabular}{c c c c c c c}
$(n,m)$ & \multicolumn{6}{c}{$T_c$ ($^\circ$C)} \\
\cline{2-7}
 & 10 & 25 & 45 & 60 & 80 & 90 \\
\hline
(2,2) & 21.1 & 14.0 & 5.6  & 0.2  & \(\times\) & \(\times\) \\
(3,2) & 20.7 & 16.7 & 11.9 & 8.7  & 4.7  & \(\times\) \\
(3,3) & 32.5 & 24.7 & 15.4 & 8.3  & \(\times\) & \(\times\) \\
(4,2) & 20.5 & 17.9 & 14.5 & 12.1 & 4.7  & \(\times\) \\
(4,3) & 32.9 & 27.7 & 20.6 & 15.5 & 7.5  & \(\times\) \\
(4,4) & 42.9 & 34.1 & 23.3 & 15.9 & 9.6  & \(\times\) \\
(5,2) & 19.4 & 19.1 & 16.1 & 6.6  & 3.8  & \(\times\) \\
(5,3) & 31.0 & 27.3 & 22.2 & 15.3 & 7.6  & \(\times\) \\
(5,4) & 41.3 & 36.3 & 30.0 & 14.6 & 9.1  & 0.6  \\
(5,5) & 52.3 & 42.8 & 30.8 & 22.4 & 10.5 & 1.6  \\
(6,2) & 17.4 & 17.0 & 15.3 & 14.0 & 3.2  & \(\times\) \\
(6,3) & 30.3 & 27.3 & 23.1 & 19.9 & 6.3  & 0.0  \\
(6,4) & 40.9 & 36.5 & 30.3 & 25.5 & 9.4  & 0.8  \\
(6,5) & 49.0 & 43.1 & 35.6 & 29.9 & 11.7 & 2.6  \\
(6,6) & 60.8 & 50.7 & 37.8 & 28.6 & 12.5 & 3.3  \\
(7,2) & 17.2 & 16.2 & 14.2 & 12.8 & 2.7  & \(\times\) \\
(7,3) & 29.5 & 27.9 & 23.8 & 20.7 & 5.4  & 0.2  \\
(7,4) & 38.6 & 35.1 & 30.0 & 25.5 & 8.1  & 0.9  \\
(7,5) & 48.0 & 43.7 & 38.0 & 28.8 & 10.8 & 2.6  \\
(7,6) & 56.1 & 49.3 & 40.7 & 31.7 & 12.3 & 3.2  \\
(7,7) & 66.2 & 57.3 & 43.5 & 31.7 & 12.3 & 3.2  \\
\end{tabular}
\end{ruledtabular}
\end{table}

Taken together with the water model, our results bridge microscopic mechanisms and operational thermodynamic constraints, offering both explanatory power and predictive universality.

\textit{In summary.---} 
Our resource-theory-based, dynamics-independent framework provides a unifying perspective on the elusive ME, addressing key limitations of earlier approaches. Unlike recent classical and quantum master-equation studies that depend on model-specific features such as spectral structure, finely tuned mode overlaps, or Liouvillian exceptional points, our formulation makes no such assumptions. Instead, it clarifies the sporadic and often contradictory experimental reports by identifying when relaxation crossovers are strictly forbidden or allowed, yielding operationally testable ``Mpemba windows'' for initial conditions. Because the framework is independent of any particular dynamical law, its predictions apply equally to Markovian and non-Markovian processes, a generality unattainable by model-bound analyses. Overall, it establishes a broadly applicable operational criterion for anomalous relaxation, capturing the essence of the ME while ensuring both universality and experimental relevance.

\textit{Acknowledgments.}---O.P. and M.H.Y. express their gratitude to Bilimler K\"{o}y\"{u} in Fo\c{ç}a, where a part of this work was conducted. O.P. and D.C.A. are grateful to Alexssandre de Oliveira Junior and Nicole Yunger Halpern for useful suggestions and extensive discussions.

%

\clearpage
\onecolumngrid  

\section*{Supplemental Material}

\setcounter{secnumdepth}{3}
\setcounter{tocdepth}{2} 

\renewcommand{\thesection}{S\arabic{section}}
\renewcommand{\thesubsection}{S\arabic{section}.\arabic{subsection}}
\renewcommand{\theequation}{S\arabic{equation}}
\renewcommand{\thefigure}{S\arabic{figure}}
\renewcommand{\thetable}{S\Roman{table}}

\setcounter{section}{0}
\setcounter{subsection}{0}
\setcounter{equation}{0}
\setcounter{figure}{0}
\setcounter{table}{0}

\section{Resource theory under thermal operations} \label{sec:SM:RTA}

Resource theories provide a general framework for modeling physical systems in which states possess a natural ordering relative to a chosen reference state. It can be viewed as a generalization of the second law of thermodynamics: states are ordered according to their distance from the reference under given operational constraints (commonly referred to as accessibility in the resource-theory literature), and this ordering dictates whether one state can be converted into another.

Within this viewpoint, a resource theory is specified by three ingredients that operationalize the “distance from a reference” described above: (i) the set of allowed (free) operations encoding the constraints; (ii) the free states, taken as references because they can be prepared at no cost (and are fixed points of the allowed operations); and (iii) the resource states, i.e., all remaining states. Given two states \(\hat\rho\) and \(\hat\sigma\), we say that \(\hat\sigma\) is \emph{accessible} from \(\hat\rho\) if some allowed operation maps \(\hat\rho\) to \(\hat\sigma\). The collection of such accessibility relations induces a preorder on states—reflexive and transitive but generally not total—so some pairs are incomparable. This induced ordering will be characterized here via thermo-majorization for energy-diagonal states under thermal operations.

\subsection{Thermo-majorization Preorder} \label{subsec:SM:TM}

When modeling a system in contact with a large heat bath, the relevant resource theory is that of  \emph{athermality}. Here, the reference (free) states are thermal states at the bath temperature \(T_b\), and the canonical choice of allowed operations is \emph{thermal operations}, defined by
\begin{equation}
    \hat{\rho} \longrightarrow \mathcal{T}(\hat{\rho}) = \operatorname{tr}_{B'}[\hat{U}(\hat{\rho}\otimes \hat{\rho}_{\rm th}(T_b,\hat{H}_B))\hat{U}^\dagger],
    \label{eq:TO-def}
\end{equation}  
where \(B\) and \(B'\) denote the bath before and after the interaction, \(\hat{H}_B\) corresponds to the bath Hamiltonian, and \(\hat{U}\) is a global energy-conserving unitary, i.e., \([\hat{U},\,\hat{H}+\hat{H}_B]=0\) such that \(\hat{H} = \sum_j E_j  |E_j\rangle \langle E_j|\) represents the self-Hamiltonian of the system.

For states diagonal in the energy basis \(\{|E_j\rangle\}\), convertibility under thermal operations is characterized by \emph{thermo-majorization}~\cite{FundLimNano2013,SecLaws2015,MeadRuch1976,AMead1977}. To apply it, we first find the Gibbs-weighted order of energy level populations at the bath temperature \(T_b\), called the \(\beta_b\)-ordering, for each state separately. Let \(\hat{\rho}=\mathrm{diag}(\vec p)\) and denote the (unnormalized) Gibbs weights as \(\mathrm{e}^{-\beta_b E_j}\), where \(\beta_b = 1/(k_B T_b)\). \(\beta_b\)-ordering of \(\hat{\rho}\) is determined by the permutation \(\vec{\pi}(\vec p)=\big(\pi_1(\vec p),\ldots,\pi_d(\vec p)\big)\) that arranges the likelihood ratios \(p_j/\mathrm{e}^{-\beta_b E_j}\) in non-increasing order, i.e.
\begin{equation}
\frac{p_{\pi_i(\vec p)}}{\mathrm{e}^{-\beta_b E_{\pi_i(\vec p)}}}
\;\ge\;
\frac{p_{\pi_{i+1}(\vec p)}}{\mathrm{e}^{-\beta_b E_{\pi_{i+1}(\vec p)}}},
\qquad i=1,\dots,d-1.
\label{eq:beta-order}
\end{equation}
Then, a piecewise-linear path called \emph{thermo-majorization curve} is constructed from the cumulative points
\begin{equation}
\mathcal{P}_{i}(\vec p)=\Bigg(\sum_{x=1}^{i}\mathrm{e}^{-\beta_b E_{\pi_x(\vec p)}},\; \sum_{x=1}^{i}p_{\pi_x(\vec p)}\Bigg),
\label{eq:TM-curve}
\end{equation}
with the conventions \(\mathcal{P}_{0}(\vec p)=(0,0)\) and \(\mathcal{P}_{d}(\vec p)=\big(Z_b(\hat H),1\big)\), where \(Z_b(\hat H)=\sum_{j=1}^d \mathrm{e}^{-\beta_b E_j}\).

For another diagonal state \(\hat\sigma=\mathrm{diag}(\vec q)\), the same construction yields \(\vec \pi(\vec q)\) and \(\mathcal{P}_i(\vec q)\); the \(\beta_b\)-orderings may differ, and both curves are considered over the common abscissa interval \([0, Z_b(\hat H)]\). We say that \(\hat\rho\) thermo-majorizes \(\hat\sigma\) (denoted \(\hat\rho\succ_{T_b}\hat\sigma\)) when the curve of \(\hat{\rho}\) lies entirely above (or on) that of \(\hat\sigma\) throughout this interval.

When the system Hamiltonian \(\hat{H}\) remains fixed, these curves can be constructed directly from the system’s energy levels. If \(\hat{H}\) changes during the process, thermo-majorization still applies, but one must include an explicit “clock” system to consistently account for the work cost of Hamiltonian changes.

Thermo-majorization can be interpreted as a \emph{relative majorization} with respect to the Gibbs state at the bath temperature, thereby providing an operational criterion for determining which state is farther from equilibrium in the presence of a bath at \(T_b\). Moreover, the family of R\'{e}nyi free energies  
\begin{equation}
    F_\alpha[\hat{\rho},T_b] 
= k_B T_b \, D_\alpha(\hat{\rho}\|\hat{\rho}_{\rm th}(T_b))
- k_B T_b \log Z_b(\hat{H}),
\label{eq:FR}
\end{equation}
where \(D_\alpha\) is the R\'{e}nyi divergence (see \eqref{eq:SM:RenyiD}), constitute a complete set of monotones for this ordering.
In particular, the conditions 
\(F_\alpha[\hat{\rho},T_b] \geq F_\alpha[\hat{\sigma},T_b]\) 
for all \(\alpha \in [0,\infty]\) is a \emph{necessary} condition 
for the existence of a thermal operation mapping \(\hat{\rho}\) to \(\hat{\sigma}\).  
However, these conditions alone are not sufficient: full convertibility requires thermo-majorization, which captures correlations among populations beyond what any finite set of scalar monotones can represent.

\subsection{Proof of Theorem 1} \label{subsec:SM:Thm1}

\textit{Theorem 1.} (No-go for Gibbs states)---
Fix a bath at \(T_b\) and consider a system with Hamiltonian \(\hat H\). For Gibbs states \(\hat\rho_{\mathrm{th}}(T_h,\hat H)\) and \(\hat\rho_{\mathrm{th}}(T_c,\hat H)\) with \(T_h>T_c>T_b\), (continuous) thermo-majorization guarantees that
\(\hat\rho_{\mathrm{th}}(T_c,\hat H)\) is always closer to equilibrium than
\(\hat\rho_{\mathrm{th}}(T_h,\hat H)\).
Thus, colder-but-farther configurations resulting in Mpemba effects are impossible within the product Gibbs family under (Markovian) thermal operations. This formalizes the thermodynamic intuition that, within the Gibbs family, hotter states are always farther from equilibrium than colder ones.

\textit{Proof.}---
Let \(\hat{\rho}_{\rm th}(T,\hat{H})=\sum_{i=1}^d p_i(T) |E_i\rangle\langle E_i|\) with 
\(p_i(T)=e^{-\beta E_i}/Z_\beta\), \(\beta=1/(k_B T)\), and \(Z_\beta=\sum_{j=1}^d e^{-\beta E_j}\).
Thermo-majorization at bath temperature \(T_b\) is defined with respect to the \(\beta_b\)-ordering, induced by the scaled populations \(p_i(T)/e^{-\beta_b E_i}\).

\emph{Step 1 (common ordering).}
For any Gibbs state, the \(\beta_b\)-ordering is independent of \(T\) and coincides with the \emph{descending} energy order. Indeed,
\begin{equation}
    \frac{p_i(T)}{e^{-\beta_b E_i}} 
= \frac{e^{-\beta E_i}/Z_\beta}{e^{-\beta_b E_i}} 
= \frac{e^{(\beta_b-\beta)E_i}}{Z_\beta}.
\label{eq:scaled-pop}
\end{equation}
If \(T\ge T_b\) (i.e. \(\beta\le \beta_b\)), the factor \(e^{(\beta_b-\beta)E_i}\) is (weakly) increasing in \(E_i\).
Consequently, the \(\beta_b\)-ordering sorts levels by non-increasing energy.
Let \(\vec{\pi}\) denote the permutation with \(E_{\pi_1}\ge \cdots \ge E_{\pi_d}\).

\emph{Step 2 (pointwise dominance at breakpoints).}  
Define the thermo-majorization breakpoints  
\(X_i=\sum_{x=1}^{i} e^{-\beta_b E_{\pi_x}}\) and \(Y_i(T)=\sum_{x=1}^{i} p_{\pi_x}(T)\),
with the conventions \(X_0=0,\;Y_0(T)=0\) and \(X_d=Z_b(\hat H),\;Y_d(T)=1\).  
Since the breakpoints \(\{X_i\}\) are identical for all \(T\ge T_b\), the thermo-majorization curves differ only in their ordinates \(\{Y_i(T)\}\). Let  
\begin{equation}  
\delta_i = \frac{e^{-\beta_h E_{\pi_i}}}{Z_h} - \frac{e^{-\beta_c E_{\pi_i}}}{Z_c}  
\end{equation}  
denote the population difference at level \(\pi_i\).  
The difference in the ordinates of the thermo-majorization curves at the \(i\)-th breakpoint is then  
\begin{equation}  
\Delta_i = \sum_{j=1}^i \delta_j = Y_i(T_h)-Y_i(T_c).  
\end{equation}

Suppose there exists an index \(m\) such that \(\delta_i>0\) for all \(i\le m\), while \(\delta_i<0\) for all \(i>m\).  
This structure is a direct consequence of the Gibbs distribution: as the temperature increases, the highest-energy levels become more populated in the hotter state than in the colder one, while the lower-energy levels become less populated. As a result, the population differences are strictly positive for the top \(m\) levels and strictly negative thereafter. This implies \(\Delta_i>0\) for all \(i\le m\). For the remaining \(d-m\) levels one has \(\delta_i<0\). By log-convexity of Gibbs weights, the sequence \(\{\delta_i\}\) decreases monotonically with \(i\).  
Therefore, once \(\delta_i\) becomes negative, its absolute value can only increase, so the increments of \(\Delta_i\) are strictly decreasing. Since both curves terminate at the common endpoint \((X_d,1)\), any crossing after \(m\) is impossible. If the hot curve were to fall below the cold one at some \(i>m\), the subsequent increments \(\delta_j<0\) with increasing magnitude would keep driving it further down, preventing it from ever reaching the common endpoint again. Hence, once it drops below, it can never recover to meet the cold curve at the final point, which contradicts the assumption that both end together. Hence the inequality \(\Delta_i>0\) for \(i\le m\) propagates consistently to all breakpoints, and the thermo-majorization curve of \(\hat\rho_{\rm th}(T_h,\hat H)\) lies entirely above that of \(\hat\rho_{\rm th}(T_c,\hat H)\).

\emph{Conclusion.}
Therefore, \(\hat\rho_{\rm th}(T_h,\hat H)\succ_{T_b}\hat\rho_{\rm th}(T_c,\hat H)\) for all \(T_h>T_c>T_b\).
Equivalently, within the Gibbs family, hotter states are strictly farther from equilibrium than colder ones. Thus, colder-but-farther configurations cannot arise under thermal operations. \(\square\)

\emph{Remark (continuity).}
Since the \(\beta_b\)-ordering is identical for all relevant states, continuous thermo-majorization (Markovian restriction) coincides with the standard notion of thermo-majorization in this setting.

\subsection{Role of Coherence} \label{subsec:SM:coh}

The discussion above focused on states diagonal in the eigenbasis of the system Hamiltonian \(\hat{H}\), for which thermo-majorization provides the necessary and sufficient condition for convertibility. By definition, however, the presence of quantum coherence drives the system further away from equilibrium, representing a nonequilibrium resource not visible in populations alone. This raises the question of how coherence transforms under thermal operations. Do the off-diagonal elements of the density matrix decay monotonically, independently of the populations, or can they influence relaxation in tandem with the energy distributions? To address this, the system’s state is decomposed in the energy eigenbasis as   
\begin{equation}
    \hat{\rho} = \sum_{n,m} \rho_{nm} |E_n\rangle \langle E_m| = \sum_{\omega \in \Omega} \hat{\rho}^{(\omega)},
    \label{eq:mode-decomp}
\end{equation}  
where \(\Omega\) is the Bohr spectrum of transition frequencies \(\omega = E_n-E_m\) between energy levels. Each operator \(\hat{\rho}^{(\omega)}\) is referred to as a \(\omega\)-mode of coherence~\cite{2014_PRA_ModesOfCoh,2015_PRX_ModesOfCoh}, explicitly given by  
\begin{equation}
    \hat{\rho}^{(\omega)} \equiv \sum_{E_n-E_m=\omega} \rho_{nm} |E_n\rangle \langle E_m|.
    \label{eq:modes}
\end{equation}

Thermal operations act independently on each mode; i.e.,  
\begin{equation}
    \mathcal{T}(\hat{\rho}^{(\omega)}) = \mathcal{T}(\hat{\rho})^{(\omega)} = \hat{\sigma}^{(\omega)}.
\end{equation}

To connect coherence with population-based thermo-majorization, one can diagonalize the density matrix within each degenerate energy subspace by means of a corresponding unitary transformation \(\mathcal{U}_\alpha\) that preserves energy and is therefore thermodynamically free. Each such \(\mathcal{U}_\alpha\) acts only inside the degenerate eigenspace labeled by \(\alpha\), leading to a block-diagonal form
\[
\hat{\rho} \xrightarrow{\{\mathcal{U}_\alpha\}} \hat{\rho}^* = \bigoplus_\alpha \hat{\rho}^*_\alpha .
\]  
If all coherence resides in the zero mode, the collection \(\mathcal{U}_\alpha\) fully diagonalizes \(\hat{\rho}\), allowing thermo-majorization to be applied directly to \(\hat{\rho}^*\): 
\[
\hat{\rho} \xrightarrow{\{\mathcal{U}_\alpha\}} \hat{\rho}^* \xrightarrow{\mathcal{T}^*} \hat{\sigma} .
\] 
Here, \(\hat{\rho}\) and \(\hat{\rho}^*\) are equivalent under the thermo-majorization pre-order. In other words, zero-mode of coherence does not circumvent the thermo-majorization constraints: it can be “translated” into an equivalent incoherent state \(\hat{\rho}^*\) while preserving the same ordering properties. In the more general case where non-zero modes are present, only the zero-mode coherences are removed by \(\{\mathcal{U}_\alpha\}\). Thermo-majorization can then be applied solely to this zero-mode component, while coherences associated with other modes remain and decay independently under thermal operations, without influencing the population-based ordering.
\begin{figure}[t]
    \includegraphics[width=.9\textwidth]{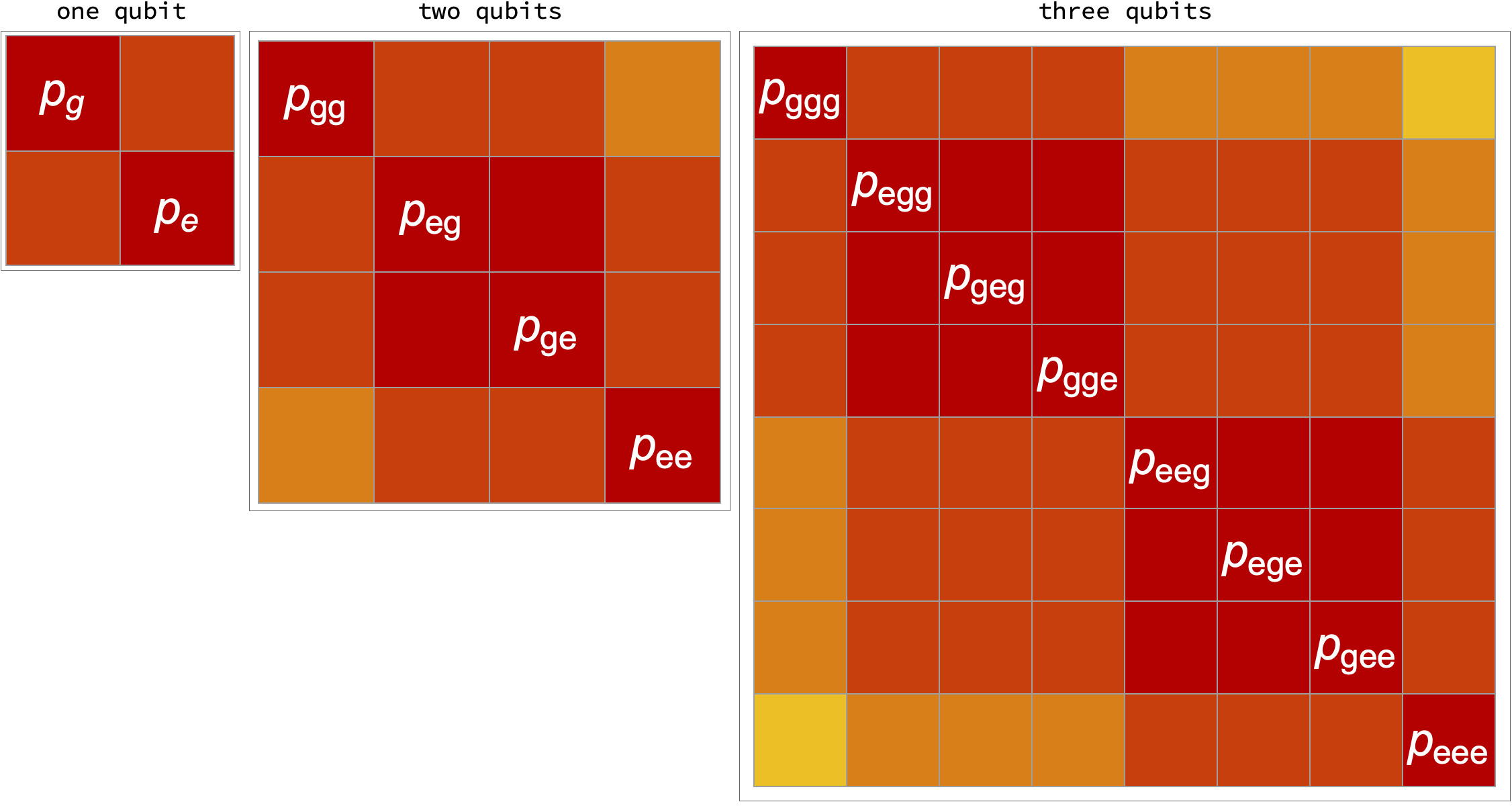}
    \caption{\justifying
    Visualization of coherence modes for one, two, and three identical qubits. Labels on the diagonals indicate the energy levels of the qubits; for example, \(p_{eg}\) denotes the population of the state with one qubit in \(|E_1\rangle=|g\rangle\) and the other in \(|E_2\rangle=|e\rangle\). Different colors highlight distinct coherence modes: red entries represent zeroth-order coherences, which contribute to thermalization. Progressively lighter shades, shifting toward yellow, correspond to higher-order coherences—dark orange for first-order, lighter orange for second-order, and yellow for third-order coherences.}
    \label{fig:cohcolored}
\end{figure}

This structure can be illustrated using the examples in Fig.~\ref{fig:cohcolored}.  
For a single qubit with energy eigenstates \(|E_1\rangle=|g\rangle\) and \(|E_2\rangle=|e\rangle\), the density matrix has off-diagonal terms belonging to the mode \(\omega=E_e-E_g\) (orange block). The diagonal populations \(\rho_{gg}=p_g\) and \(\rho_{ee}=p_e\) (red block) evolve independently of these coherences under thermal operations.

For two qubits, three distinct coherence modes arise: \(\omega\in\{0,\,E_e-E_g,\,2(E_e-E_g)\}\), corresponding respectively to the red, dark orange, and lighter orange blocks in Fig.~\ref{fig:cohcolored}. The zeroth-order block (red) contains not only the diagonal populations such as \(\rho_{eg,eg}=p_{eg}\) and \(\rho_{ge,ge}=p_{ge}\), but also off-diagonal elements like \(\rho_{eg,ge}\) and \(\rho_{ge,eg}\). These evolve jointly, and block-diagonalizing this sector leaves the thermo-majorization properties unchanged.

For three qubits, there are two distinct zeroth-order blocks: one in the single-excitation sector and one in the double-excitation sector. Each can be diagonalized independently without altering the ordering. Higher-order blocks evolve independently during thermal relaxation, with each color in Fig.~\ref{fig:cohcolored} denoting a different order of coherence. 

Thus, while coherence certainly constitutes a form of athermality, only the zero-mode sector can be fully translated into an equivalent incoherent contribution within thermo-majorization through such an energy-preserving block diagonalization. Higher-order coherence modes evolve independently under thermal operations and, although dynamically relevant, do not modify the population-based thermo-majorization ordering.

\subsection{Continuous Thermo-Majorization} \label{subsec:SM:contTM}

Thermo-majorization fully characterizes convertibility of block-diagonal states under general thermal operations—including non-Markovian ones—by determining whether one state is operationally farther from, or closer to, equilibrium than another. Markovian relaxation, however, imposes an additional continuity constraint: it demands a continuous trajectory of admissible thermal operations connecting the initial and final states. Within this setting, \textit{continuous thermo-majorization} provides the appropriate partial order for diagonal states in the energy basis~\cite{ContThermo2022}.  

The fundamental building blocks of continuous thermo-majorization are the \emph{partial level thermalizations} (PLTs). These are thermal operations that act non-trivially only on two energy levels and admit a simple master equation description:
\begin{equation}
    \frac{dp_i}{dt}=\frac{1}{\tau}\!\left[\frac{\gamma_i}{\gamma_i+\gamma_j}(p_i+p_j)-p_i\right], 
\qquad 
\frac{dp_j}{dt}=-\frac{dp_i}{dt}, \label{eq:plt-ode}
\end{equation}
where \(p_i\) and \(p_j\) are the two populations involved,  \(\gamma_i\) and \(\gamma_j\) are the corresponding equilibrium weights of the Gibbs distribution, and \(\tau\) is a relaxation parameter. The solution takes the form
\begin{equation}
    \vec{p}(t) = T^{i,j}(\lambda_t) \, \vec{p}(0), \qquad 
\lambda_t=1-e^{-t/\tau},
\label{eq:plt-solution}
\end{equation}
with transition matrix
\begin{equation}
    T^{i,j}(\lambda_t) =
\begin{bmatrix}
(1-\lambda_t)+\lambda_t\frac{\gamma_i}{\gamma_i+\gamma_j} & \lambda_t \frac{\gamma_i}{\gamma_i+\gamma_j} \\
\lambda_t \frac{\gamma_j}{\gamma_i+\gamma_j} & (1-\lambda_t)+\lambda_t\frac{\gamma_j}{\gamma_i+\gamma_j}
\end{bmatrix} \otimes \mathbb{I}_{\setminus (j,k)}.
\end{equation}

\begin{figure}[t]
    \centering
    \includegraphics[width=0.7\textwidth]{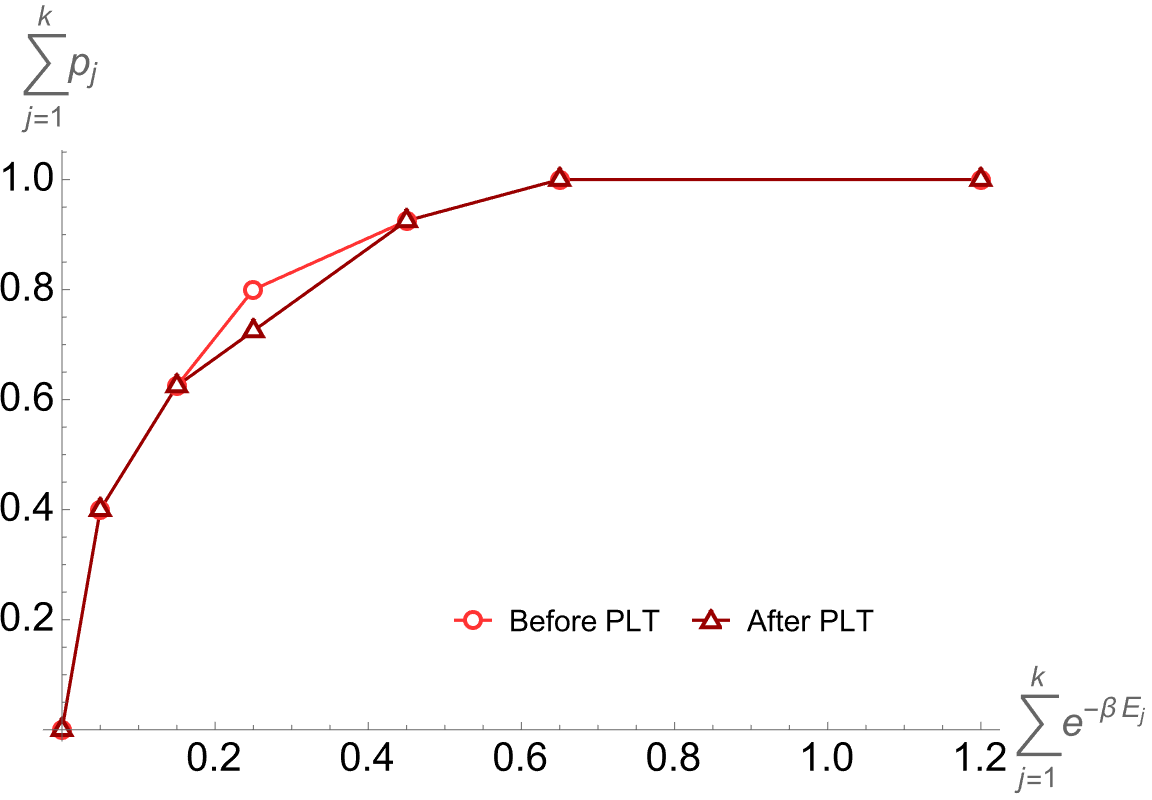}
    \caption{\justifying Effect of a partial level thermalization (PLT) on the thermo-majorization curve. The lighter curve is mapped to the darker one after applying \(T^{3,4}(\lambda=1)\), which equalizes the slopes in the relevant segment, corresponding to a full two-level thermalization.}
    \label{fig:plt}
\end{figure}

The parameter \(\lambda\) ranges from 0 to 1, interpolating between the identity operation and the full equilibration of the two levels. Figure~\ref{fig:plt} illustrates how applying \(T^{i,j}(\lambda=1)\) equalizes the slopes of the corresponding segments in the thermo-majorization curve.

The significance of PLTs lies in the fact that any Markovian thermal operation can be decomposed into a finite sequence of such steps. Specifically, for any Markovian transition \[\hat{\rho}\xrightarrow{\text{MTO}}\hat{\sigma},\] there exists a sequence of indices \((i_k,j_k)\) and parameters \(\lambda_k\) such that
\begin{equation}
    \hat{\sigma} = T^{i_f,j_f}(\lambda_f)\cdots T^{i_1,j_1}(\lambda_1)\hat{\rho}.
\end{equation}

In this case, we say that \(\hat{\rho}\) \emph{continuously thermo-majorizes} \(\hat{\sigma}\). A constructive protocol uses the \(\beta_b\)-ordering. Consider two population vectors \(\vec{p}\) and \(\vec{q}\). We enumerate canonical sequences of \(\beta_b\)-orderings \(\{\vec \pi^k\}_{k=1}^N\) that begin with the ordering of \(\vec{p}\) (\(\vec \pi^1=\vec \pi(\vec{p})\)) and end with that of \(\vec{q}\) (\(\vec \pi^N=\vec \pi(\vec{q})\)), where successive permutations differ only by swapping two adjacent elements. From each such sequence, we construct
\begin{equation}
    \vec{f} := \prod_{k=1}^{N-1} T^{i_k,j_k}(1)\,\vec{p},
    \label{eq:contTM-check}
\end{equation}
where \((i_k,j_k)\) denote the swapped indices. Then \(\vec{p}\) continuously thermo-majorizes \(\vec{q}\) if and only if at least one such \(\vec{f}\) thermo-majorizes \(\vec{q}\).

Importantly, whenever \(\vec{p}\) and \(\vec{q}\) share the same \(\beta_b\)-ordering, continuous thermo-majorization coincides with standard thermo-majorization. In this case, the Markovian restriction does not impose any additional constraints, and the two notions coincide.

\section{Correlation-enabled Mpemba effects}
\label{sec:SM:CorrME}

In this section, we investigate realizations of the Mpemba effect that are enabled by correlations. To this end, we use the same set of initial states in both the generalized and the original scenarios. The first initial state is a product Gibbs preparation. For the sake of simplification, we consider two qubits at temperature \(T_x\):
\begin{equation}
  \hat{\rho}_P(2,T_x) = \hat{\rho}_{\mathrm{th}}(T_x,\hat{H}_q)\otimes \hat{\rho}_{\mathrm{th}}(T_x,\hat{H}_q) = \mathrm{diag}(\vec{p}_P(T_x)),
\end{equation}
with \(\hat{H}_q\) the single-qubit Hamiltonian  with eigenvalues \(E_1<E_2\) and the diagonal vector
\begin{equation}
  \vec{p}_P(T_x) = \frac{1}{Z_x^2}\big(e^{-2\beta_x E_1},\, e^{-\beta_x(E_1+E_2)},\, e^{-\beta_x(E_1+E_2)},\, e^{-2\beta_x E_2}\big),
\end{equation}
where the partition function is \(Z_x = e^{-\beta_x E_1}+e^{-\beta_x E_2}\) and \(\beta_x=1/(k_B T_x)\).

The second initial state is a correlated preparation at temperature \(T_x\)
\begin{equation}
    \hat{\rho}_C(2,T_x) = \frac{1}{Z_x}\Big(e^{-\beta_x E_1}\,|E_1E_1\rangle \langle E_1E_1| + e^{-\beta_x E_2}\,|E_2E_2\rangle \langle E_2E_2|\Big) = \mathrm{diag}(\vec{p}_C(T_x)),
\end{equation}
with the diagonal vector
\begin{equation}
  \vec{p}_C(T_x) = \frac{1}{Z_x}\big(e^{-\beta_x E_1},\,0,\,0,\,e^{-\beta_x E_2}\big).
\end{equation}

Each marginal of both initial states corresponds to a Gibbs state \(\hat{\rho}_{\mathrm{th}}(T_x,\hat{H}_q)\), yet the joint statistics are correlated in the second case as defined in the main text.

\subsection{Generalized Mpemba Effect}
\label{subsec:SM:GenME}

We first consider the case where both qubits interact weakly with local heat baths at temperature \(T_b\). The total Hamiltonian is 
\begin{equation}
  \hat{H} = \hat{H}_q\otimes \mathbb{I} + \mathbb{I}\otimes \hat{H}_q,
\end{equation}
with \(\hat{H}_q = E_1 |E_1\rangle\langle E_1| + E_2 |E_2\rangle\langle E_2|\). The joint state \(\hat{\rho}(t)\) evolves under the local GKSL master equation
\begin{equation}
  \frac{d\hat{\rho}}{dt} = -\frac{i}{\hbar}[\hat{H},\hat{\rho}] + \mathcal{D}_1[\hat{\rho}] + \mathcal{D}_2[\hat{\rho}],
  \label{eq:SM-localME}
\end{equation}
where the dissipators are defined by
\begin{equation}
  \mathcal{D}_j[\hat{\rho}] = \gamma_j^\downarrow \Big(\sigma_-^{(j)} \hat{\rho} \, \sigma_+^{(j)} - \tfrac12\big\{\sigma_+^{(j)} \sigma_-^{(j)},\hat{\rho}\big\}\Big) 
  + \gamma_j^\uparrow \Big(\sigma_+^{(j)} \hat{\rho} \,\sigma_-^{(j)} - \tfrac12\big\{\sigma_-^{(j)} \sigma_+^{(j)},\hat{\rho}\big\}\Big),
\end{equation}
with \(\sigma_-^{(1)}=|E_1\rangle\langle E_2|\otimes \mathbb{I}\),  
\(\sigma_-^{(2)}=\mathbb{I}\otimes |E_1\rangle\langle E_2|\),  
and \(\sigma_+^{(j)}=(\sigma_-^{(j)})^\dagger\). Here \(\gamma_j^\downarrow\) denotes the emission (downward) rate and \(\gamma_j^\uparrow\) the absorption (upward) rate, satisfying detailed balance
\begin{equation}
  \gamma_j^\uparrow/\gamma_j^\downarrow = e^{-\beta_b \Delta}, 
  \qquad \Delta=E_2-E_1,\quad \beta_b=1/(k_B T_b).
\end{equation}

Because each dissipator accounts for all qubit-level transitions and the rates obey detailed balance, every initial state relaxes to the Gibbs state at the bath temperature \(T_b\). Moreover, initial states diagonal in the energy basis, such as \(\hat{\rho}_P(2,T_x)\) and \(\hat{\rho}_C(2,T_x)\), remain incoherent throughout the evolution. That is, the initial probability vectors \(\vec{p}_P(T_x)\) and \(\vec{p}_C(T_x)\) evolve toward the stationary distribution
\begin{equation}
  \vec{\gamma} \equiv \vec{p}_P(T_b) = \frac{1}{Z_b^2}\big(e^{-2\beta_b E_1},\, e^{-\beta_b(E_1+E_2)},\, e^{-\beta_b(E_1+E_2)},\, e^{-2\beta_b E_2}\big),
\end{equation}
with partition function \(Z_b = e^{-\beta_b E_1}+e^{-\beta_b E_2}\) and \(\beta_b=1/(k_B T_b)\), and during this evolution, each qubit can always be assigned an instantaneous effective temperature determined by its marginal population ratio.

In the literature, the occurrence of Mpemba-like effects is typically explained through the overlap of the initial state with the slowest decaying mode(s) of the Liouvillian generator. Within this framework, a state with a larger overlap is expected to relax more slowly. If the same criterion were adopted here, the slowest mode would be \(- \kappa \equiv - (\gamma^\uparrow + \gamma^\downarrow) \) for the diagonal states with two degenerate eigenoperators \(\{\hat{l}_1, \hat{l}_2\}\) and \(\{\hat{r}_1, \hat{r}_2\}\) such that \(\mathcal{G}[\hat r_j] = - \kappa \hat r_j\) and \(\mathcal{G}^\dagger[\hat l_j] = - \kappa \hat l_j\), where \( \frac{d\hat{\rho}}{dt} = \mathcal{G} \hat{\rho}\). Following Ref.~\cite{moroder2024thermodynamics}, we quantify the overlap with the slowest-decaying Liouvillian modes as \(\mathcal{O}_j[\hat \rho (t_0)] = |{\rm tr}[\hat{l}_j \hat \rho (t_0)]|\). For instance, the product Gibbs preparation has overlaps
\begin{equation}
    \mathcal{O}_1[\hat \rho_P(2,T_x)] = \frac{(e^{\beta_x E_2 + \beta_b E_1} - e^{\beta_x E_1 + \beta_b E_2})^2}{(e^{\beta_x E_1}+e^{\beta_x E_2})^2 (e^{\beta_b E_1}+e^{\beta_b E_2})^2} \, , \qquad \mathcal{O}_2[\hat \rho_P(2,T_x)] = \frac{e^{\beta_x E_1} - e^{\beta_x E_2 + \beta_b E_1 - \beta_b E_2}}{e^{\beta_x E_1}+e^{\beta_x E_2}} \, ;
\end{equation}
whereas the correlated preparation has the overlaps
\begin{equation}
    \mathcal{O}_1[\hat \rho_C(2,T_x)] = \frac{e^{\beta_x E_2 + 2 \beta_b E_1} + e^{\beta_x E_1 + 2 \beta_b E_2}}{(e^{\beta_x E_1}+e^{\beta_x E_2})(e^{\beta_b E_1}+e^{\beta_b E_2})^2} \, , \qquad \mathcal{O}_2[\hat \rho_C(2,T_x)] = \frac{e^{\beta_x E_1} - e^{\beta_x E_2 + \beta_b E_1 - \beta_b E_2}}{e^{\beta_x E_1}+e^{\beta_x E_2}} \, ;
\end{equation}
which results in the prediction that the correlated preparation \(\hat \rho_C(2,T_c)\) may thermalize more slowly than the product preparation \(\hat \rho_P(2,T_h)\) although \(T_h > T_c\). For example, when the energy spectrum is taken as \(E_1=0\) eV and \(E_2 = 0.05\) eV and the local thermalization processes are described by \(T_b = 0^\circ\) C and  \(\gamma_1^\uparrow = \gamma_2^\uparrow = 10^0 \, \text{s}^{-1}\), it turns out that \(\mathcal{O}_j[\hat \rho_C(2,60^\circ \text{C}]=\{0.129, 0.047\}\) and \(\mathcal{O}_j[\hat \rho_P(2,136.7^\circ \text{C}]=\{0.008, 0.099\}\). However, our explicit analysis reveals the opposite: although \(\hat \rho_C(2,T_c)\) starts further away from equilibrium, it approaches equilibrium faster than the hotter product state. This demonstrates that the conventional mode-overlap mechanism, though frequently invoked in the literature, is insufficient to account for the effect observed in our setting. The observed behavior instead highlights the genuinely correlation-enabled nature of the Mpemba phenomenon in this model.
\begin{figure}[t]
  \centering
  \begin{subfigure}[t]{0.48\linewidth}
    \centering
    \includegraphics[width=\linewidth]{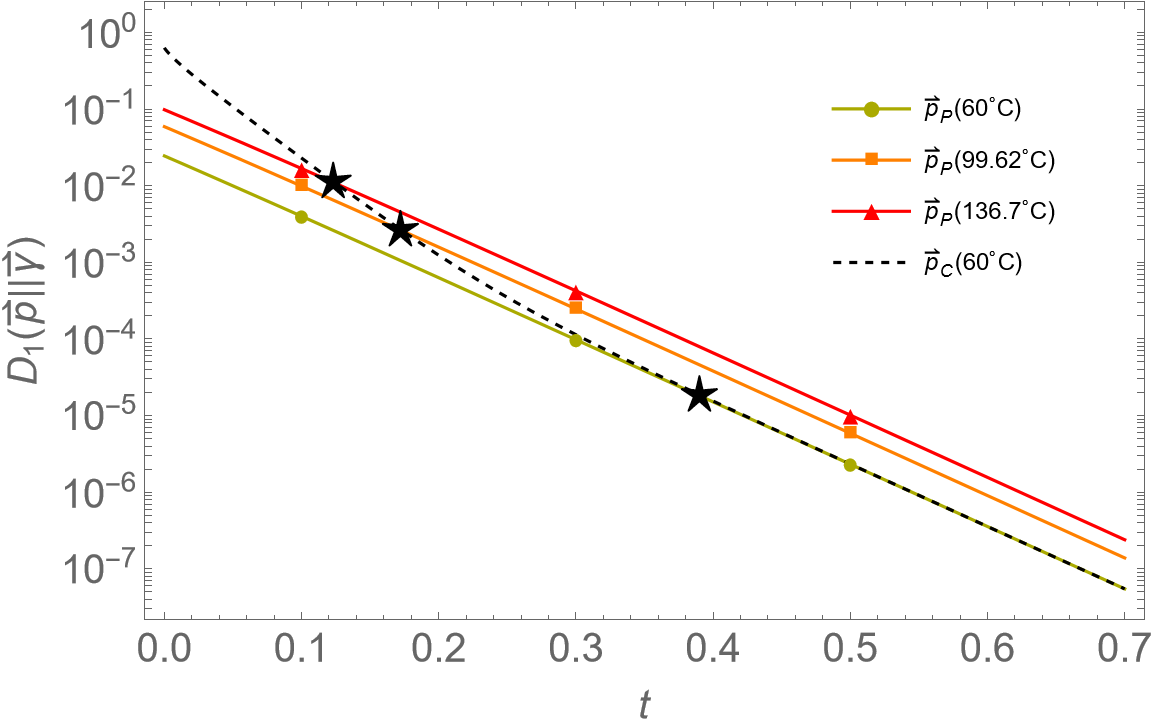}
  \end{subfigure}
  \hfill
  \begin{subfigure}[t]{0.48\linewidth}
    \centering
    \includegraphics[width=\linewidth]{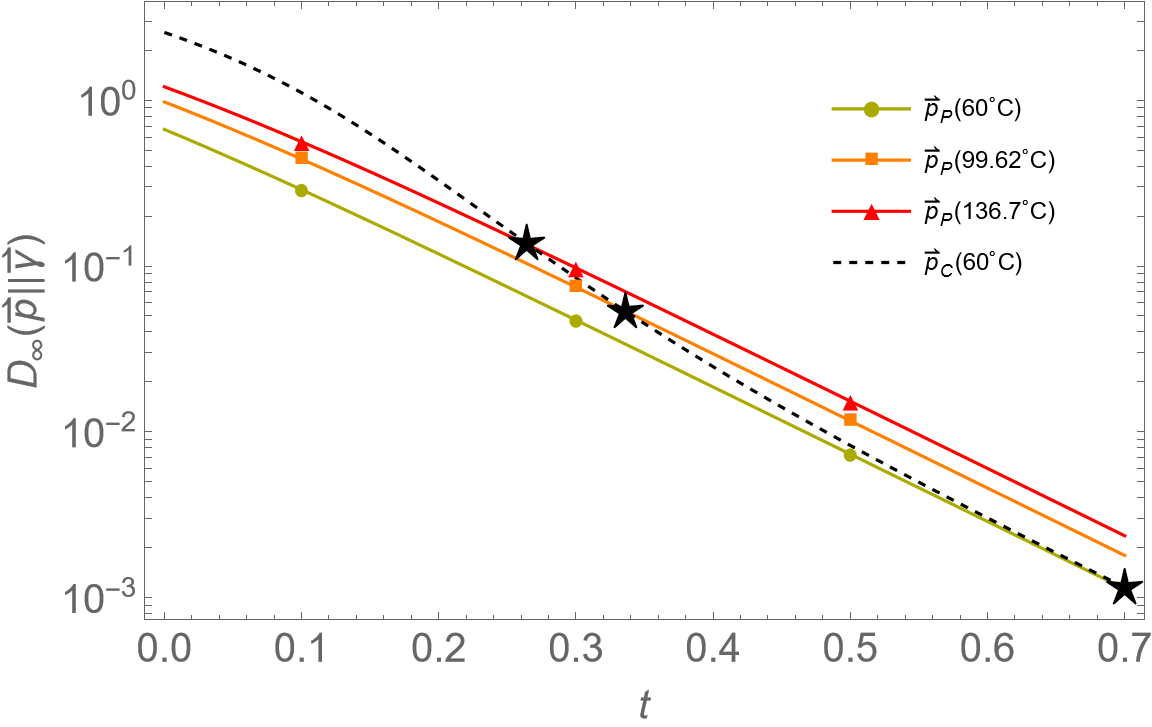}
  \end{subfigure}
  \caption{\justifying
  Comparison of \(D_1(\vec{p}(t)\|\vec{\gamma})\) and \(D_\infty(\vec{p}(t)\|\vec{\gamma})\) for different initial preparations with \(E_1=0\) eV, \(E_2 = 0.05\) eV, \(T_b = 0^\circ\)C, and \(\gamma_1^\uparrow = \gamma_2^\uparrow = 1 \,\text{s}^{-1}\). Both metrics demonstrate the generalized Mpemba effect under local master-equation dynamics. (left panel) Crossovers in the Kullback--Leibler divergence trajectories (\(\star\)) reveal the effect. (right panel) Order reversals in the maximum divergence trajectories (\(\star\)) confirm its robustness, enabled by correlations.}
  \label{fig:SM-D-local}
\end{figure}

\begin{figure}[t]
  \centering \includegraphics[width=0.48\linewidth]{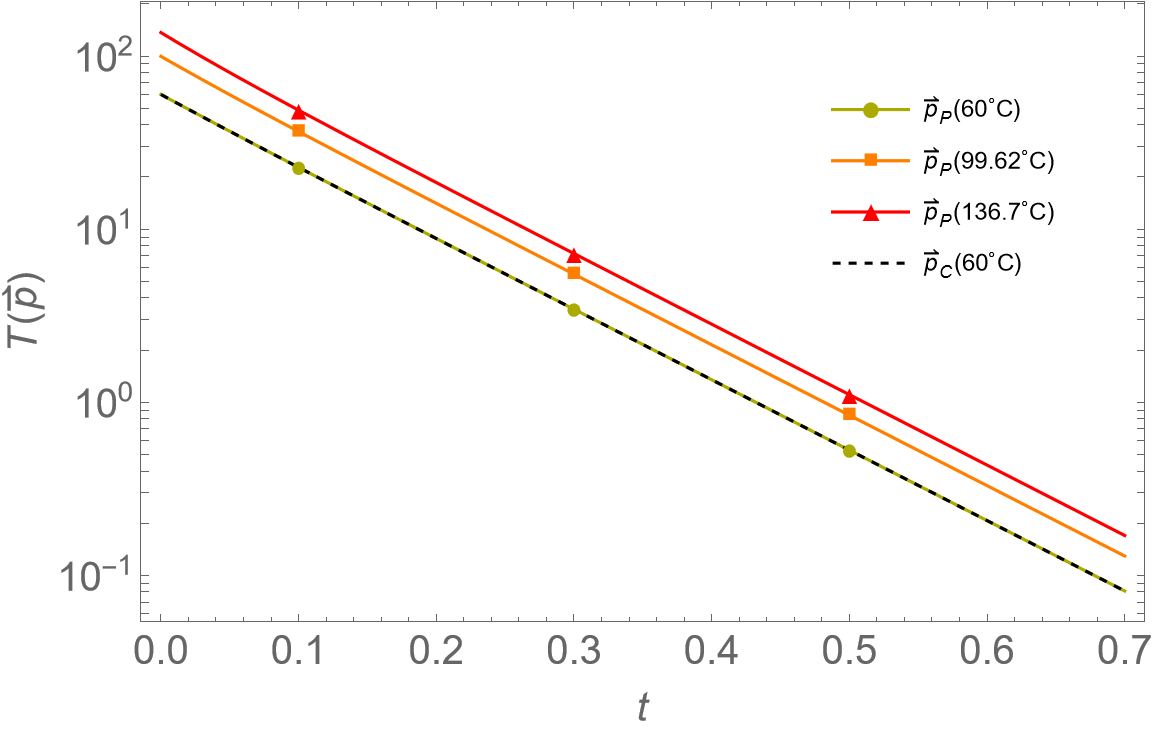}
  \caption{\justifying
  Effective temperature \(T(\vec{p}(t))\) for the same set of preparations as in Fig.~\ref{fig:SM-D-local}. The hotter preparations remain hotter along the trajectories, yet a generalized Mpemba effect emerges when assessed via athermality monotones.}
  \label{fig:SM-Teff-local}
\end{figure}

To characterize relaxation, we monitor the R\'{e}nyi divergences
\begin{equation}\label{eq:SM:RenyiD}
  D_\alpha(\vec{p}(t)\,\|\,\vec{\gamma}) = \frac{1}{\alpha -1} \log \sum_i p_i^\alpha \gamma_i^{1-\alpha}, \qquad \alpha\in (0,1) \cup (1,\infty),
\end{equation}
and an effective temperature \(T(\vec{p}(t))\) obtained by fitting single-qubit marginals to \(\hat{\rho}_{\mathrm{th}}(T,\hat{H}_q)\). Using the initial states defined above, we observe order reversals in the distance-to-equilibrium curves. When the energy spectrum is taken as \(E_1=0\) eV and \(E_2 = 0.05\) eV and the local thermalization processes are described by \(T_b = 0^\circ\) C and  \(\gamma_1^\uparrow = \gamma_2^\uparrow = 10^0 \, \text{s}^{-1}\), a colder-but-farther preparation \(\hat{\rho}_C(2,60^\circ \text{C})\) can approach equilibrium faster than the hotter-but-closer preparations \(\hat{\rho}_P(2,T_h)\) with \(T_h \in [60^\circ\text{C},136.7^\circ\text{C}]\) (see Fig.~\ref{fig:SM-D-local}). In contrast, the local temperature curves do not exhibit any crossing as shown in Fig.~\ref{fig:SM-Teff-local}. This constitutes a generalized Mpemba effect, enabled by correlations in the initial state. Thus, while our qubit example does not realize the conventional mode-overlap mechanism, it serves as a minimal and analytically transparent model of correlation-enabled Mpemba effects.

\subsection{Original Mpemba Effect}
\label{subsec:SM:OrigME}
\begin{figure}[t]
  \centering
  \begin{subfigure}[t]{0.48\linewidth}
    \centering
    \includegraphics[width=\linewidth]{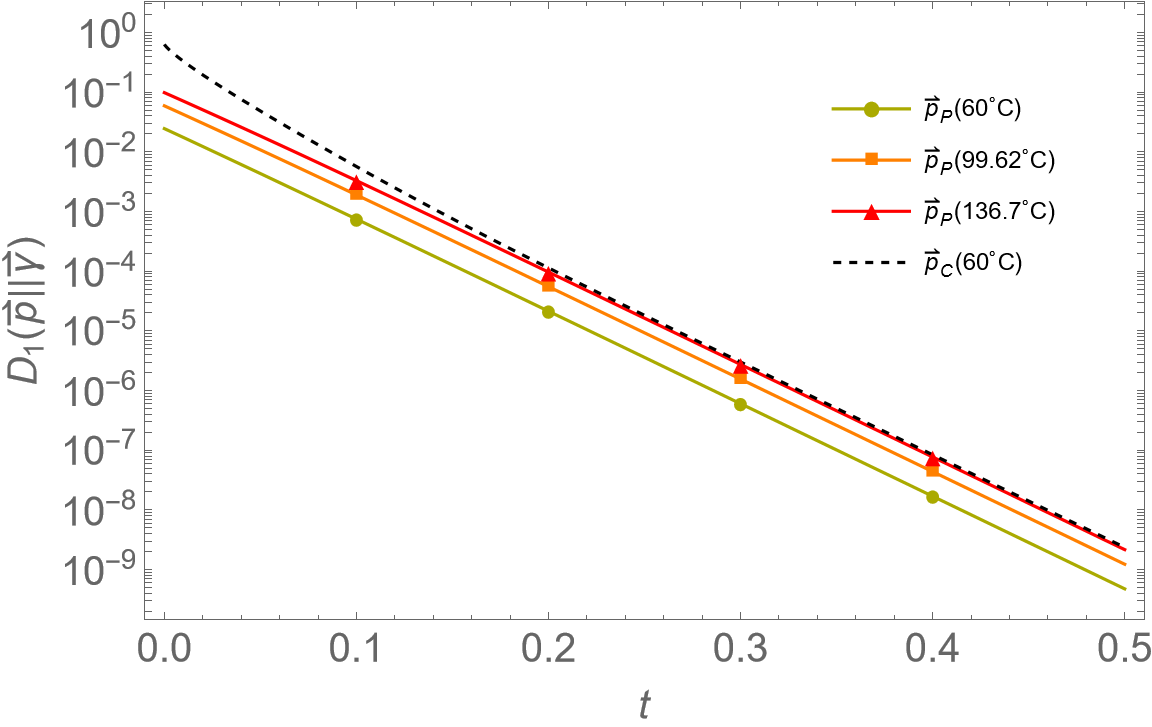}
  \end{subfigure}
  \hfill
  \begin{subfigure}[t]{0.48\linewidth}
    \centering
    \includegraphics[width=\linewidth]{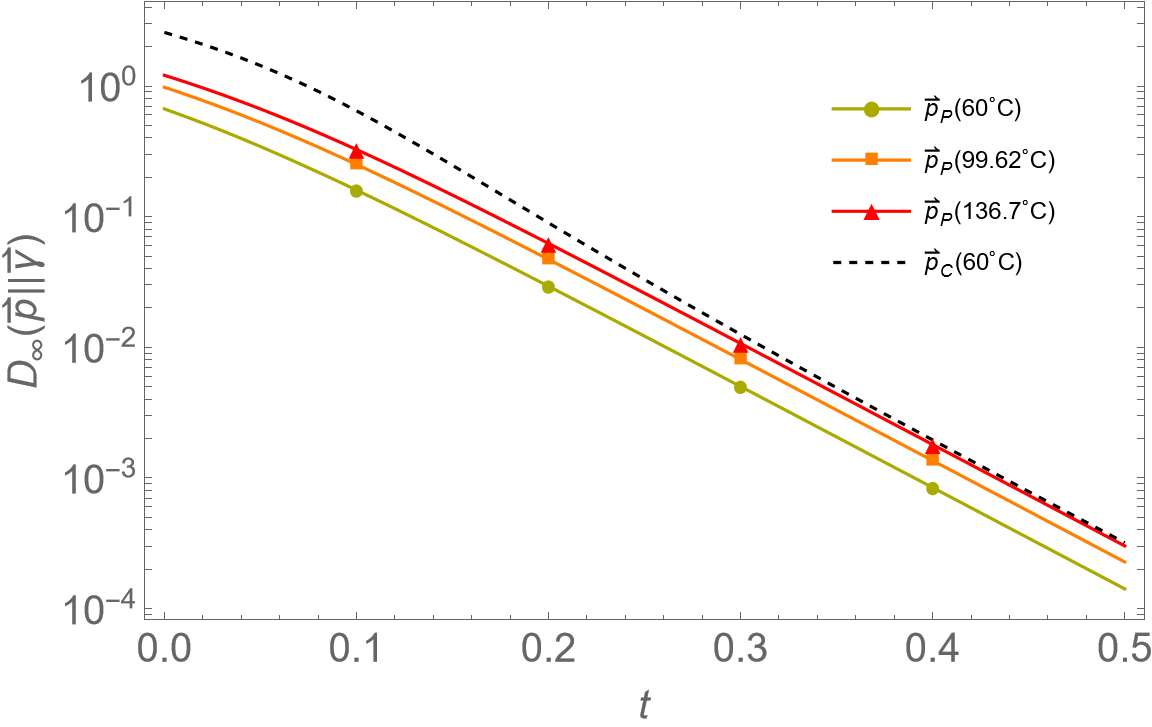}
  \end{subfigure}
  \caption{\justifying
  Comparison of \(D_1(\vec{p}(t)\|\vec{\gamma})\) and \(D_\infty(\vec{p}(t)\|\vec{\gamma})\) for different initial preparations with \(E_1=0\) eV, \(E_2=0.05\) eV, \(T_b=0^\circ\)C, and \(\gamma_1^\uparrow=\gamma_2^\uparrow=\gamma_{1|1}^\uparrow=\gamma_{2|1}^\uparrow=2\gamma_{1|2}^\uparrow=2\gamma_{2|2}^\uparrow=1\,\text{s}^{-1}\). (left panel) No crossovers appear in the Kullback--Leibler divergence trajectories. (right panel) The absence of order reversals in the maximum divergence trajectories confirms that no generalized Mpemba effect occurs under conditional channels.}
  \label{fig:SM-D-cond}
\end{figure}
\begin{figure}[t]
  \centering \includegraphics[width=0.48\linewidth]{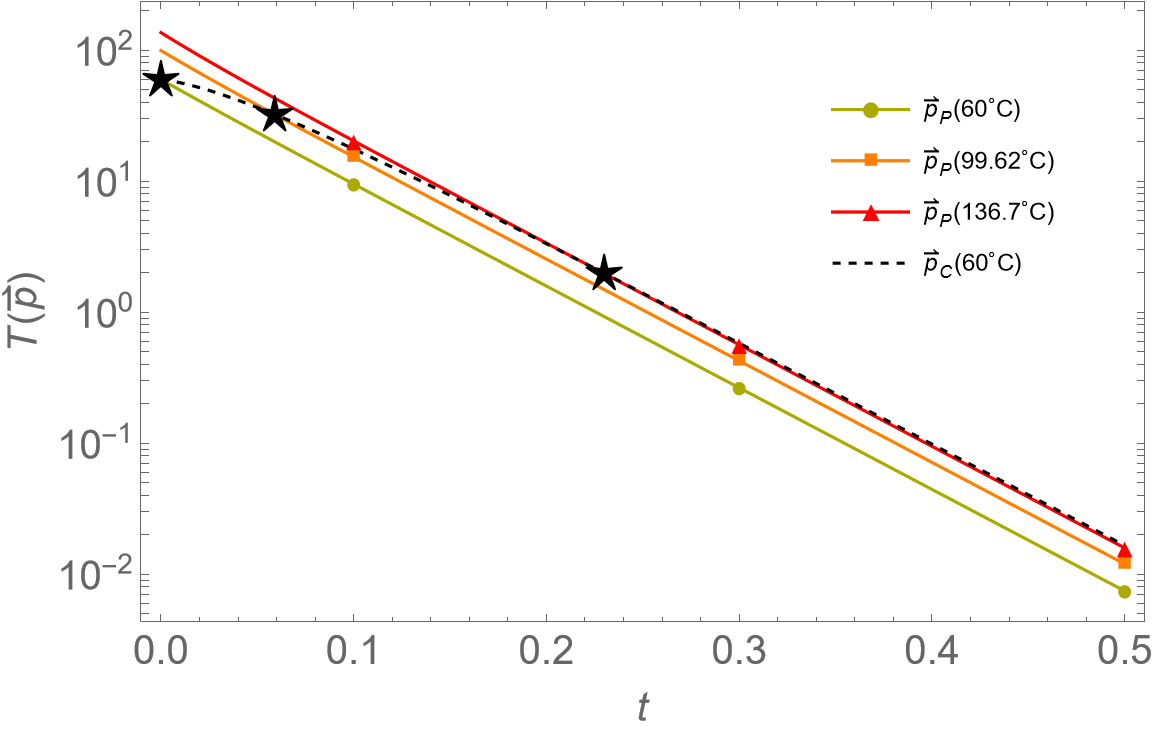}
  \caption{\justifying
  Effective temperature \(T(\vec{p}(t))\) for the same set of preparations as in Fig.~\ref{fig:SM-D-cond}. Crossovers between hotter-but-closer and colder-but-farther trajectories (\(\star\)) reveal an original-type Mpemba effect, enabled by initial correlations.}
  \label{fig:SM-Teff-cond}
\end{figure}

We now extend the local thermalization model of Eq.~\ref{eq:SM-localME} by introducing four additional dissipation channels. Each additional dissipator implements a state-dependent transition of one qubit, conditioned on the energy configuration of the other:
\begin{equation}
  \mathcal{D}_{j|k}[\hat{\rho}] = \gamma_{j|k}^\downarrow \Big(\sigma_-^{(j|k)} \hat{\rho} \, \sigma_+^{(j|k)} - \tfrac12\big\{\sigma_+^{(j|k)} \sigma_-^{(j|k)},\hat{\rho}\big\}\Big) 
  + \gamma_{j|k}^\uparrow \Big(\sigma_+^{(j|k)} \hat{\rho} \,\sigma_-^{(j|k)} - \tfrac12\big\{\sigma_-^{(j|k)} \sigma_+^{(j)},\hat{\rho}\big\}\Big),
\end{equation}
with \(\sigma_-^{(1|k)}=|E_1\rangle\langle E_2|\otimes |E_k\rangle\langle E_k|\),  
\(\sigma_-^{(2|k)}=|E_k\rangle\langle E_k|\otimes |E_1\rangle\langle E_2|\),  
and \(\sigma_+^{(j|k)}=(\sigma_-^{(j|k)})^\dagger\). The new rates also satisfy detailed balance,
\begin{equation}
  \gamma_{j|k}^\uparrow/\gamma_{j|k}^\downarrow = e^{-\beta_b \Delta},
\end{equation}
which guarantees that that \(\hat \rho_P(2, T_b)\) is the fixed point of the dynamics.

As in the local case, restricting to diagonal initial preparations \(\hat \rho_P(2, T_x)\) and \(\hat \rho_C(2, T_x)\) ensures that no coherence is generated, allowing us to consistently assign the same instantaneous effective temperature to both qubits. In this setting, the distance-to-equilibrium measures do not display any crossings (Fig.~\ref{fig:SM-D-cond}). However, when relaxation is monitored via effective local temperatures, we observe a crossover between hotter-but-closer and colder-but-farther trajectories (Fig.~\ref{fig:SM-Teff-cond}). This constitutes an \emph{original-type Mpemba effect}, enabled by correlations in the initial state and manifested directly in the temperature dynamics, rather than in generalized athermality monotones.

Our analysis reveals two distinct correlation-enabled manifestations of the Mpemba effect. In the local bath scenario, correlations induce a \emph{generalized Mpemba effect}, where colder-but-farther states relax faster than hotter-but-closer ones when relaxation is assessed through athermality monotones, while local temperature curves remain ordered. In contrast, when conditional dissipation channels are introduced, correlations give rise to an \emph{original-type Mpemba effect}, where order reversals are directly observed in the effective temperature trajectories without any crossing in athermality monotones. Taken together, these results demonstrate that correlations both enable and diversify Mpemba-like anomalies, giving rise to qualitatively distinct behaviors that cannot be explained by the conventional mode-overlap mechanism alone.

\section{Thermodynamic limits of correlation-enabled Mpemba effects}
\label{sec:SM:TDLimits}

The explicit master-equation examples discussed in the previous section demonstrate that initial correlations can give rise to new forms of the Mpemba effect: faster relaxation of states farther from equilibrium without temperature crossings, and faster cooling of hotter systems even in the absence of monotone crossings. Although thermal relaxation is generally modeled by Gibbs-preserving maps, such dynamics do not always correspond to genuine thermal operations. Establishing their physical realizability, therefore, requires a microscopic justification. To circumvent this constraint, we adopt a resource-theoretic approach and employ thermo-majorization to characterize how correlations alter relaxation orderings and to determine the fundamental thermodynamic limits of these effects independently of any particular dynamical model.

\subsection{Mpemba Window in Qubit Systems}
\label{subsec:SM:qubits}

To illustrate how correlations expand the space of possible relaxation behaviors, we begin with the identical qubit systems analyzed in the main text. Each qubit is described by the Hamiltonian 
\[
\hat{H}_q = E_g |g\rangle \langle g| + E_e |e\rangle \langle e| .
\]
In the hot sample, all \(n\) qubits are uncorrelated and locally thermal at \(T_h\), 
\begin{equation}\label{eq:SM:rhoh4q}
  \hat{\rho}_{\rm hot} = \hat{\rho}_P(n, T_h) = \hat{\rho}_{\rm th}(T_h, \hat{H}_q)^{\otimes n} .
\end{equation}

\subsubsection{Classically Correlated Local Thermal States}
\label{subsec:SM:CCqubits}

The cold sample is locally thermal at \(T_c < T_h\) but contains correlations among a subset of qubits,
\begin{equation}\label{eq:SM:rhoc4q}
    \hat{\rho}_{\rm cold} = \hat{\rho}_{C}(m, T_c) \otimes \hat{\rho}_P(n-m, T_c) ,
\end{equation}
where the correlated block of size \(m\) is
\begin{equation}
    \hat{\rho}_{C}(m, T_c) = P_g(T_c)\, (|g\rangle\langle g|)^{\otimes m} + P_e(T_c)\,(|e\rangle\langle e|)^{\otimes m} , \quad P_{g/e}(T_c) = e^{-\beta_c E_{g/e}}/Z_c(\hat{H}_q) .
\end{equation}

\subparagraph{Illustrative example: two-qubit systems.}

The minimal nontrivial case \(m = n = 2\) already suffices to demonstrate correlation-enabled reversals of thermodynamic ordering. In Sec.~\ref{sec:SM:CorrME} we examined the relaxation of \(\hat{\rho}_{\rm hot} = \hat{\rho}_P(2, T_h) = \mathrm{diag}(\vec{p}_P(T_h))\) and \(\hat{\rho}_{\rm cold} = \hat{\rho}_C(2, T_c) = \mathrm{diag}(\vec{p}_C(T_c))\) using Markovian master equations. Here, we turn to a dynamics-independent formulation based on thermo-majorization, which captures all operations allowed by thermal resource theory and is not restricted to Markovian dynamics.
\begin{figure} [t]
    \centering
    \begin{subfigure}[b]{0.49\textwidth}
        \includegraphics[width=\textwidth]{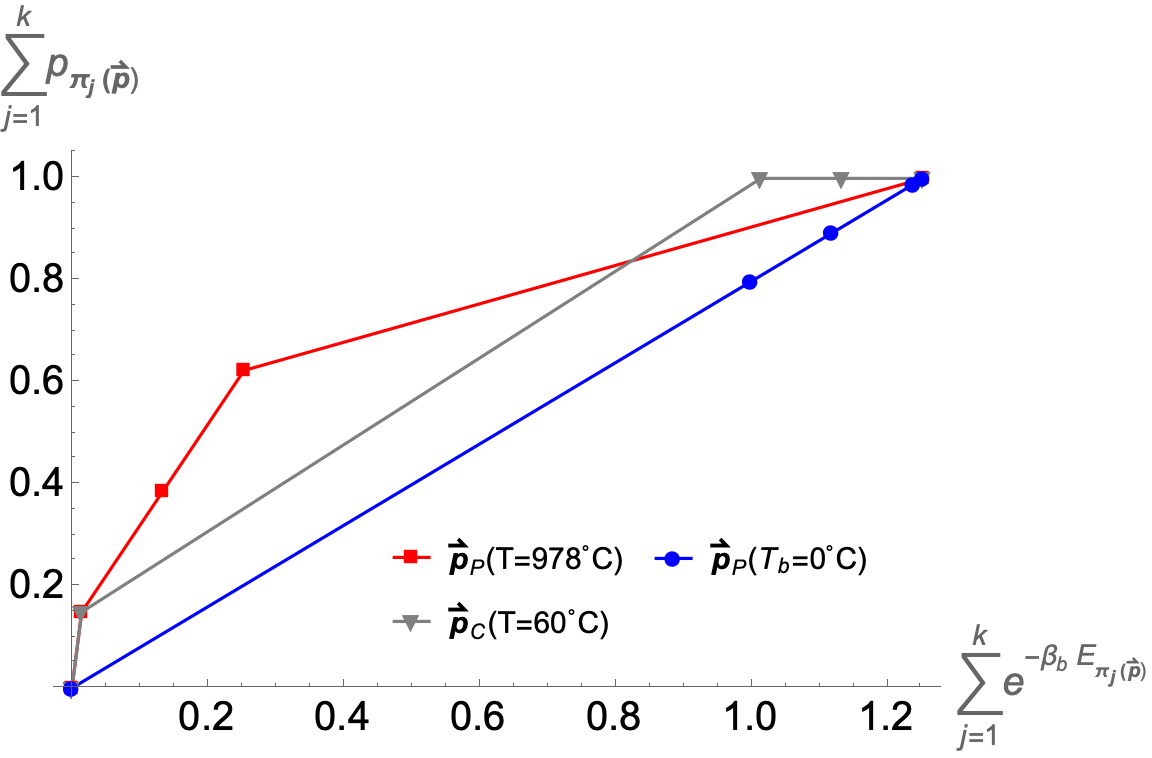}
    \end{subfigure}
        \begin{subfigure}[b]{0.49\textwidth}
        \includegraphics[width=\textwidth]{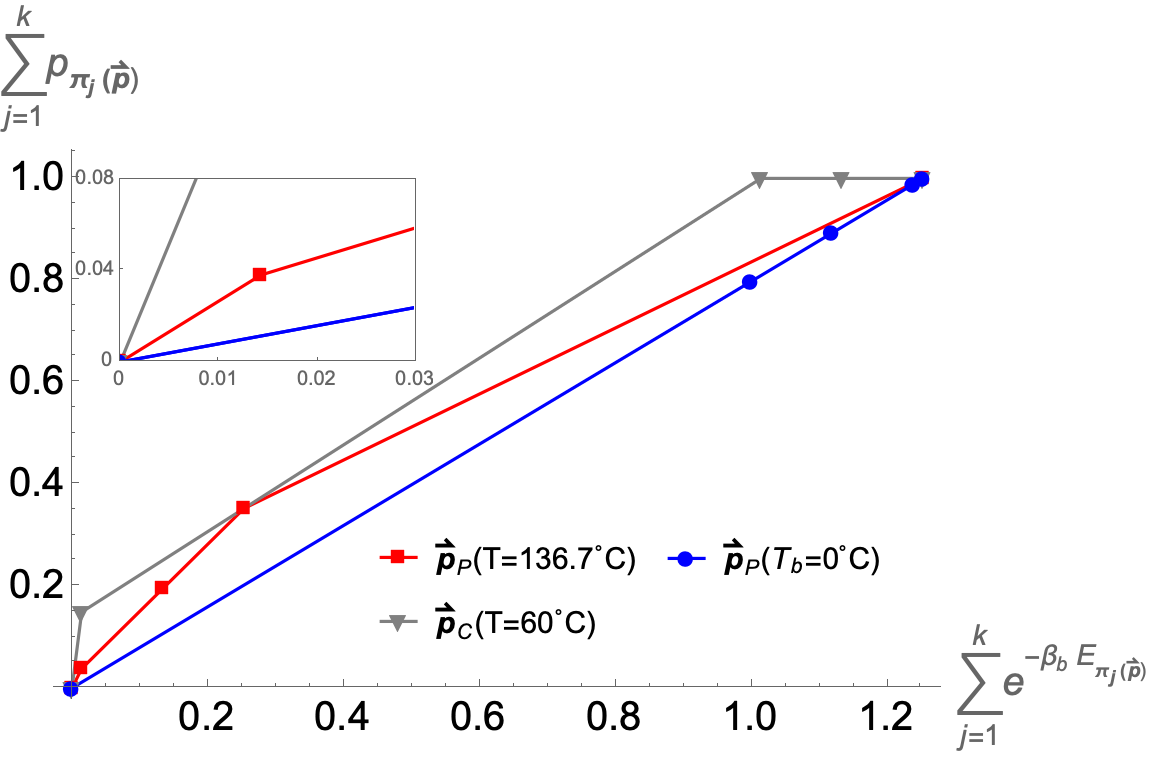}
    \end{subfigure}
    \caption{\justifying Thermo-majorization analysis for a pair of classically correlated qubits prepared at \(T_c = 60^\circ\)C. Our algorithm iteratively determines the maximum temperature \(T_h^{\max}\) for which the hot product state \(\vec{p}_P(T_h)\) remains thermo-majorized by the correlated cold state \(\vec{p}_C(T_c)\) relative to a bath at \(T_b = 0^\circ\)C. Single-qubit energies are \(E_g = 0\) and \(E_e = 0.05\) eV. (Left) First step: \(T_h = 978^\circ\) C aligns the first points of the curves, yet \(\mathcal{P}_{prod}\) subsequently lies above \(\mathcal{P}_{corr}\). (Right) Final step: allowing the third points to coincide yields \(T_h^{\max} = 136.7^\circ\)C, where the remaining segments satisfy the majorization relation. If the temperature of \(\mathcal{P}_{prod}\) is increased any further, its third point will go above \(\mathcal{P}_{corr}\), and the states become incomparable.}
    \label{fig:SM:2qC}
\end{figure}

Fixing \(\mathcal{C}(T_c)\) in Definition 3 of the main text to the correlated two-qubit state \(\hat{\rho}_{\rm corr}=\hat{\rho}_{C}(2,T_c)\), we determine the maximal temperature \(T_h^{\max}(T_c)\) for which \(\vec{p}_C(T_c) \succ_{T_b} \vec{p}_P(T_h)\). The relevant population vectors reordered by decreasing \(\beta_b\)-weighted probabilities are 
\begin{eqnarray}
    \vec{p}_P &=& (e^{-2 \beta_h E_e}, e^{-\beta_h \left(E_e + E_g\right)}, e^{-\beta_h \left(E_e + E_g\right)}, e^{-2 \beta_h E_g})/Z_P , \\
    \vec{p}_C &=& (e^{-\beta_c E_e}, e^{-\beta_c E_g}, 0, 0)/Z_C ,
\end{eqnarray}
with \(Z_P = Z_h(\hat{H}_q \otimes \mathbb{I} + \mathbb{I} \otimes \hat{H}_q)\) and \(Z_C = Z_c(\hat{H}_q)\).

The first points of the thermo-majorization curves derived from these vectors have equal horizontal components. Therefore, a necessary condition for a correlation-enabled Mpemba effect is
\begin{equation}\label{eq:SM:cond14c}
    e^{-2 \beta_h E_e}/Z_P \leq e^{-\beta_c E_e}/Z_C.
\end{equation}
Satisfaction of this inequality alone, however, is insufficient: equality merely ensures that the first segments coincide, while subsequent ones may still violate the ordering, as shown in the left panel of Fig.~\ref{fig:SM:2qC}.

Moreover, because the second and third components of \(\vec{p}_P\) are identical, the corresponding segments of the product thermo-majorization curve have equal slopes. Hence, \(\mathcal{P}_{prod}\) and \(\mathcal{P}_{corr}\) can intersect at most at the third point of the product curve. This requires the additional condition
\begin{equation}\label{eq:SM:cond24c}
    (e^{-2 \beta_h E_e}+ 2 e^{-\beta_h (E_g+E_e)})/Z_P = (e^{-\beta_c E_e}+ 2 e^{-\beta_b (E_e-E_g) - \beta_c E_g})/Z_C .
\end{equation}

The temperature \(T_h\) satisfying both constraints~\eqref{eq:SM:cond14c} and~\eqref{eq:SM:cond24c} defines the upper edge of the Mpemba window \([T_c,T_h^{\max}(T_c)]\). For example,  when the single-qubit energy levels are set to \(E_g = 0\) eV and \(E_e = 0.05\) eV, the classically correlated state \(\hat{\rho}_C\) at \(60^\circ\)C becomes farther from equilibrium than the product state \(\hat{\rho}_P\) with a temperature as high as \(136.70^\circ\)C during relaxation toward \(0^\circ\)C (see right panel in Fig.~\ref{fig:SM:2qC}). This inversion of thermal ordering directly quantifies the thermodynamic advantage conferred by correlations.

\subparagraph{Markovian limit: continuous thermo-majorization.}
\begin{figure}[b]
    \includegraphics[width=0.7\textwidth]{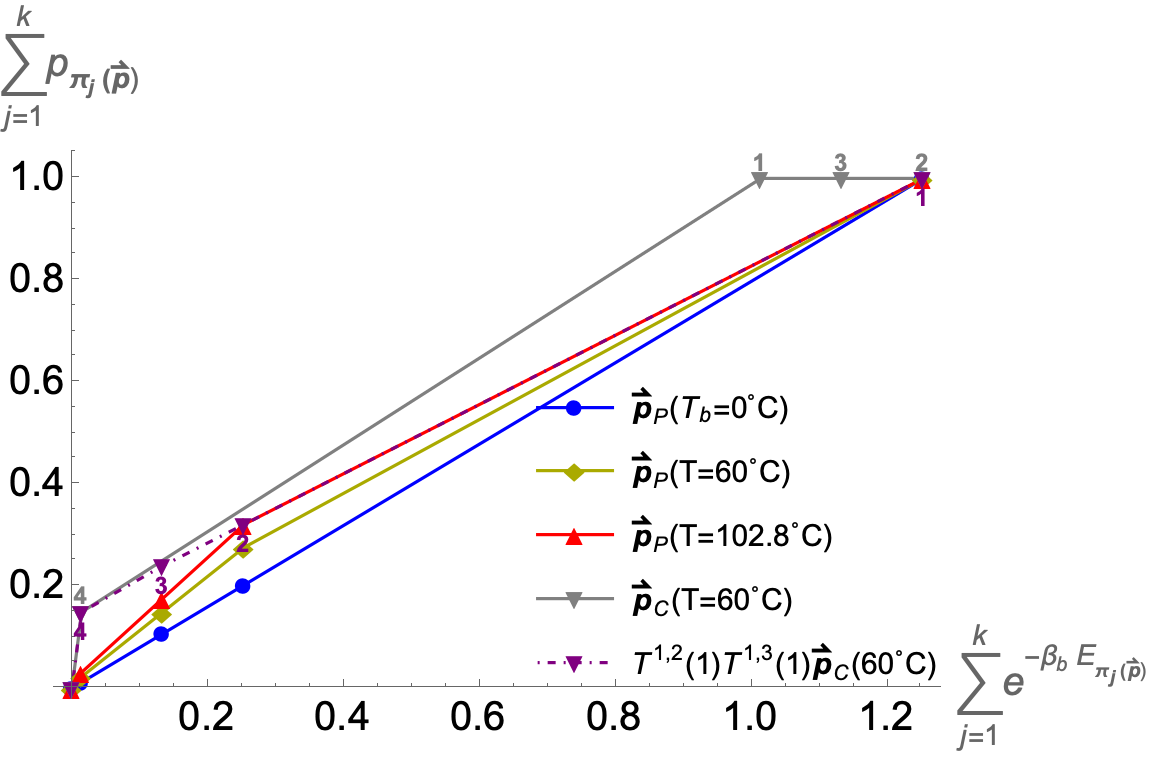}
    \caption{\justifying Continuous thermo-majorization analysis for a pair of classically correlated qubits prepared at \(T_c = 60^\circ\)C. The grey curve shows the correlated population vector \( \vec{p}_C\), while the dashed purple curve represents its image under the sequence of complete thermalizations \( T^{1,2}(1) \, T^{1,3}(1)\vec{p}_C\). The maximum-temperature product state \( \vec{p}_P (102.8^\circ \text{C})\) (red) just touches the transformed correlated curve at the point labeled 2. The initial ordering of the correlated state \(\left( 4, 1, 3, 2\right)\) matches that of the product state after these thermalizations, indicating that under Markovian thermal operations the Mpemba window narrows to \(T_h^{\max} = 102.8^\circ\)C. The single-qubit energy levels are identical to those used in Fig.~\ref{fig:SM:2qC}.}
    \label{fig:SM:2qCcont}
\end{figure}

The thermo-majorization framework discussed above identifies the range of temperatures \( [T_c, T_h^{\max}(T_c)] \) for which correlations invert the thermodynamic ordering of product and correlated states. However, this analysis assumes access to arbitrary thermal operations, which need not be physically realizable through Markovian dynamics. To determine how such dynamical constraints modify the Mpemba window, we now turn to the Markovian limit, formulated in Sec.~\ref{subsec:SM:contTM} in terms of continuous thermo-majorization.

In this limit, allowed transformations correspond to infinitesimal stochastic processes that preserve the Gibbs state at the bath temperature \(T_b\). Each elementary step acts as a thermalization between a pair of energy levels, represented by a stochastic matrix \(T^{i,j}(\lambda)\) with coupling strength \(\lambda \in [0,1]\). Successive applications of such pairwise operations generate the continuous family of thermal operations describing Markovian thermal relaxation. Unlike the non-Markovian case, where full thermo-majorization allows any monotonic reordering consistent with the majorization curves, the continuous limit restricts accessible transitions to those reachable through concatenations of pairwise thermalizations.

Figure~\ref{fig:SM:2qCcont} illustrates this constraint for the same two-qubit system analyzed in Fig.~\ref{fig:SM:2qC}. Starting from the correlated distribution \(\vec{p}_C(T_c)\) at \(60^\circ\)C, successive thermalizations \(T^{1,2}(1)\,T^{1,3}(1)\) deform the corresponding thermo-majorization curve until its \(\beta_b\)-ordering aligns with that of a product state at \(T_h^{\max}=102.8^\circ\)C. The resulting intersection defines the \emph{Markovian} Mpemba window, which is narrower than the non-Markovian one (\(T_h^{\max}=136.7^\circ\)C), indicating that Markovian thermal operations impose additional constraints on the speedup attainable via correlations.

At first glance, this conclusion seems inconsistent with the results obtained in Sec.~\ref{sec:SM:CorrME}, where the relaxation of the same correlated and product states was analyzed through Markovian master equations. Those master equations can be microscopically derived from collision models and therefore correspond to Markovian thermal operations. In that section, the departure from equilibrium was quantified using athermality measures, and the relaxation dynamics were presented in Figs.~\ref{fig:SM-D-local}--\ref{fig:SM-Teff-cond}. According to these results, the Markovian Mpemba window is not \(T_h^{\max}=102.8^\circ\)C. Instead, it extends up to \(T_h^{\max}=136.7^\circ\)C, as in the non-Markovian case. The origin of this discrepancy lies in the fact that the athermality monotones employed there provide necessary but not sufficient conditions for state convertibility. Although the curve corresponding to \(T_h=136.7^\circ\)C is not shown in Fig.~\ref{fig:SM:2qCcont}, its extrapolated position would lie above the transformed correlated curve at the point labeled \(2\), rendering the two states incomparable under Markovian thermal operations. This illustrates that monotone-based analysis, although widely used in the literature, may overestimate the operational accessibility of correlation-enabled Mpemba effects.

In other words, while correlations still enable a reversal of thermal ordering, the degree of this advantage depends on the dynamical accessibility of the corresponding thermal operations. The continuous thermo-majorization formalism therefore bridges the gap between the purely resource-theoretic and physically implementable regimes of relaxation.

\subparagraph{Multi-qubit generalization.}

The analysis can be extended to multi-qubit systems where correlations are shared among arbitrary subsets of qubits.  
We again consider hot and cold samples of the forms defined in Eqs.~\eqref{eq:SM:rhoh4q}~and~\eqref{eq:SM:rhoc4q}.  
The parameters \(n\) and \(m\) respectively denote the total number of qubits and the number of correlated ones.  
For \(m=n\), correlations are \emph{global} (all-to-all), whereas for \(m<n\), they are \emph{local} or \emph{partial} (block-correlated).

Figure~\ref{fig:multiC} of the main text summarizes the maximal temperatures \(T_h^{\max}(T_c)\) obtained from the thermo-majorization criterion between cold correlated and hot product preparations at \(T_c = 60^\circ\)C for various excitation gaps \(E_e - E_g\). Panel~(a) corresponds to globally correlated (all-to-all) systems with \(m=n\), whereas panel~(b) depicts partially correlated (block-correlated) ones where only \(m<n\) qubits share correlations. In the globally correlated case, only two collective configurations contribute, so the cold correlated curve acquires a sharp-like thermo-majorization profile: after a short on-ramp, it rises steeply toward unit height and is followed by an extended tail (using the standard Lorenz-curve terminology~\cite{gour2015resource}). The hot product curve, by contrast, spreads its population over all excitation sectors and therefore develops a broader cumulative rise across many breakpoints. As the gap increases, the leading breakpoints of both thermo-majorization curves move downward together, reflecting the suppression of the fully excited configuration. For the correlated state, which has only two nontrivial breakpoints, this lowers the onset of its sharp rise toward unit height. The product curve also starts lower at its earliest breakpoints, but its population is accumulated over many excitation sectors, so its multi-step rise can overtake the correlated curve at later breakpoints. Since the correlated curve is already flat at unit height on its tail, any loss of ordering must occur before that tail begins. Accordingly, for \(E_e = 0.06\)–\(0.07\) eV, the product curve remains below the correlated one over the common abscissa interval, whereas for \(E_e = 0.08~\mathrm{eV}\) the cumulative weight of the highest-excitation sectors becomes sufficient to push the product curve through the steep pre-tail segment, thereby closing the Mpemba window.
 \begin{figure} [t]
    \centering
    \includegraphics[width=.7\textwidth]{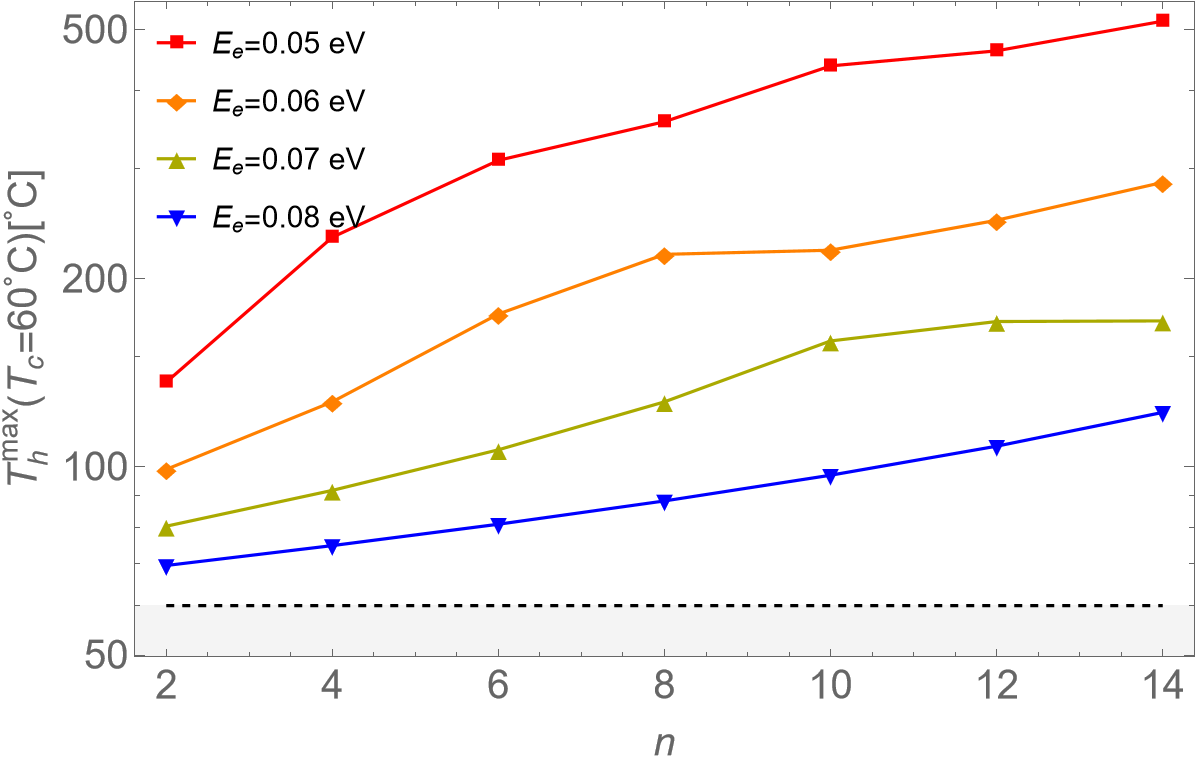}
    \caption{\justifying Maximum hot temperature \(T_h^{\max}(T_c=60^\circ\mathrm{C})\) for multi-qubit systems composed of pairwise (dimerized) correlated blocks, \(\hat{\rho}_{C}(2,T_c)^{\otimes n/2}\), obtained from the thermo-majorization criterion \(\vec{p}_C(T_c)\succ_{T_b}\vec{p}_P(T_h)\). Each curve corresponds to a different excitation gap \(E_e-E_g\), as indicated in the legend. The vertical axis is logarithmic, showing that \(T_h^{\max}\) increases nearly exponentially with system size \(n\) across all energy gaps. The persistence of the effect even for large gaps (\(E_e=0.08~\mathrm{eV}\)) contrasts with the global (all-to-all) case in Fig.~\ref{fig:multiC}~a, where the effect vanishes under similar conditions, suggesting that correlation modularity plays a key role in sustaining the Mpemba window.}
    \label{fig:SM:DimerizedMultiC}
\end{figure}
\begin{figure} [h!]
    \centering
    \includegraphics[width=.7\textwidth]{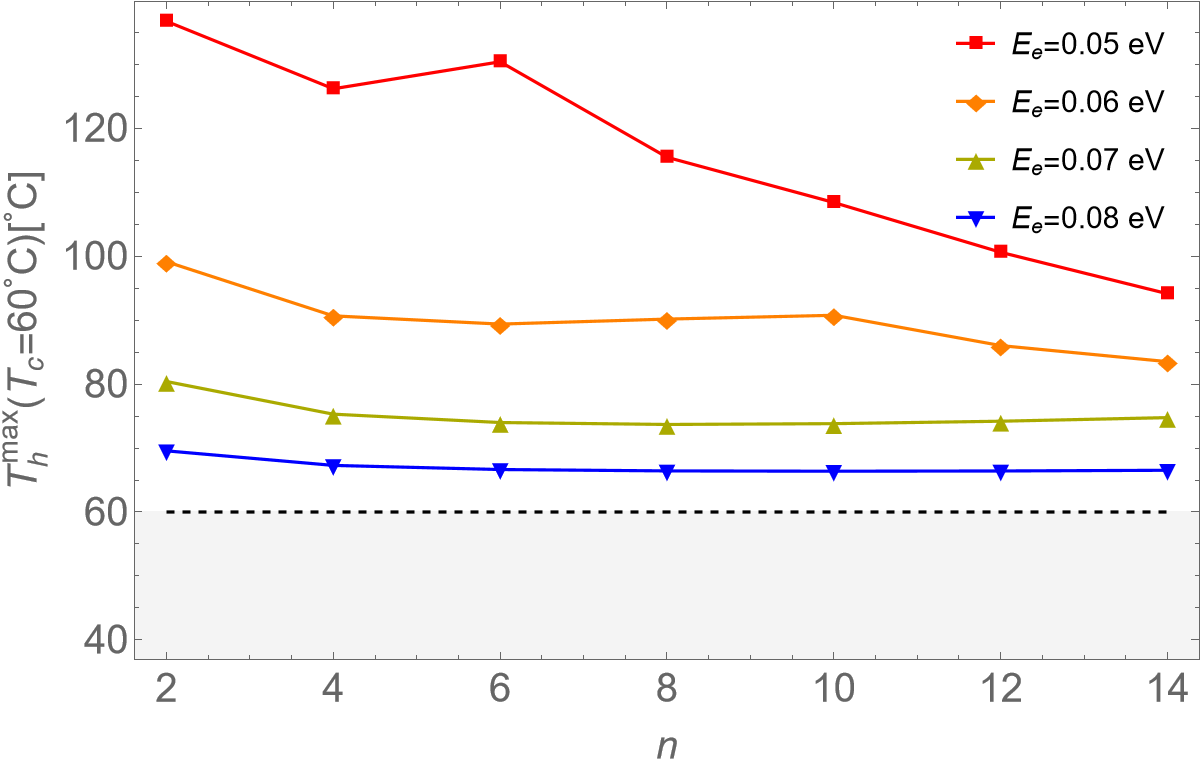}
    \caption{\justifying Maximum hot temperature \(T_h^{\max}(T_c=60^\circ\mathrm{C})\) for multi-qubit systems containing a single correlated pair, \(\hat{\rho}_{\rm cold} = \hat{\rho}_{C}(2,T_c) \otimes \hat{\rho}_{P}(n-2,T_c)\), obtained from the thermo-majorization criterion \(\vec{p}_C(T_c)\succ_{T_b}\vec{p}_P(T_h)\). Each curve corresponds to a different excitation gap \(E_e-E_g\), as indicated in the legend. Unlike the dimerized configuration in Fig.~\ref{fig:SM:DimerizedMultiC}, the Mpemba window here shows a nonmonotonic dependence on system size \(n\) that is largely insensitive to the excitation gap. This trend may stem from the uncorrelated qubits effectively diluting the population bias introduced by the single correlated pair.}
    \label{fig:SM:SingleMultiC}
\end{figure}

This same geometric rigidity also clarifies why the dependence on correlation topology is nontrivial. As indicated by the vertical lines in Fig.~\ref{fig:multiC}~a, global (all-to-all) correlations with \(m=n=11\) or \(m=n=16\) fail to produce a Mpemba inversion for sufficiently large excitation gaps. In contrast, partial (block) correlations between \(m=11\) out of \(n=16\) qubits restore the effect, as shown in Fig.~\ref{fig:multiC}~b. This distinction arises because global correlations confine the joint state to only two collective populations, producing a rigid thermo-majorization profile with minimal breakpoint structure. Block correlations, by contrast, populate multiple excitation sectors within each correlated cluster and thereby generate a more structured spectrum of cumulative populations. The additional breakpoints allow the curve to adapt more flexibly to the product-state geometry, so that ordering lost in the fully correlated case can be restored. As a result, intermediate correlation topologies, neither completely global nor entirely independent, can yield the largest thermodynamic contrast between correlated and product ensembles.

Further insight can be gained by comparing different patterns of correlation distribution. At \(T_c = 60^\circ\)C, pairwise correlations in \(\hat{\rho}_{\rm cold} = \hat{\rho}_{C}(2, T_c)^{\otimes n/2}\) cause the maximal hot temperature \(T_h^{\max}\) (and hence the Mpemba window width) to grow approximately exponentially with \(n\) over the plotted range for all excitation gaps \(E_e-E_g\) considered (see Fig.~\ref{fig:SM:DimerizedMultiC}). By contrast, when correlations are confined to a single pair, \(\hat{\rho}_{\rm cold} = \hat{\rho}_{C}(2, T_c) \otimes \hat{\rho}_{P}(n-2, T_c)\), the maximal temperature exhibits a nonmonotonic dependence on system size and remains comparatively insensitive to the excitation gap (Fig.~\ref{fig:SM:SingleMultiC}). Microscopically, the difference reflects how correlated clusters reshape the thermo-majorization curve: in the dimerized configuration, each correlated pair introduces a local population bias that is replicated across independent blocks, leading to a progressively stronger shift of \(T_h^{\max}\), whereas in the single-pair case this sharp bias is diluted by the remaining uncorrelated qubits, whose product-like populations dominate the later breakpoints. These observations show that the thermodynamic role of classical correlations depends not only on their amount but also on how they are distributed across subsystems.

Overall, the multi-qubit analysis reveals that classical correlations can facilitate colder-but-farther configurations whose thermodynamic accessibility depends sensitively on both energy spacing and correlation architecture. The observed trends point to a nontrivial interplay between local block structure and global ordering, motivating the subsequent analysis of quantum-correlated (entangled or discordant) states.

\subsubsection{Quantum Correlated Local Thermal States}
\label{subsec:SM:QCqubits}

\subparagraph{Entangled local thermal states.}

We next examine whether multipartite entanglement can enlarge the correlation-enabled Mpemba window beyond the classically correlated scenarios discussed above. As in the previous subsection, we focus first on families whose local marginals remain thermal at the same cold temperature \(T_c\), so that the comparison isolates the effect of correlation structure itself.

As a first representative family, consider cold samples of the form
\begin{equation}
  \hat{\rho}_{\rm cold}
=
\hat{\rho}_{E}(m,T_c)\otimes \hat{\rho}_{P}(n-m,T_c),
\end{equation}
where \(m\) out of \(n\) qubits share multipartite entanglement through
\begin{equation}
  \hat{\rho}_{E}(m,T_c)
=
\hat{\rho}_{C}(m,T_c)
+
\mu (|g\rangle\langle e|)^{\otimes m}
+
\mu^*(|e\rangle\langle g|)^{\otimes m},
\qquad
|\mu|\le \sqrt{P_g(T_c)P_e(T_c)}.
\end{equation}
This construction preserves the same diagonal populations as \(\hat{\rho}_{C}(m,T_c)\), while adding GHZ-like multipartite coherences between the fully grounded and fully excited configurations. In the numerical examples below, \(|\mu|\) is taken at its maximal allowed value, \(|\mu|=\sqrt{P_g(T_c)P_e(T_c)}\). Crucially, the off-diagonal terms in \(\hat{\rho}_{E}(m,T_c)\) belong entirely to nonzero modes of coherence (here \(\omega = m(E_e-E_g)\)), so the state contains no zero-mode coherence. Under thermal operations, coherences in nonzero modes evolve independently of the population sector and therefore do not modify the thermo-majorization ordering determined by the diagonal part. As a result, \(\hat{\rho}_{E}(m,T_c)\otimes \hat{\rho}_{P}(n-m,T_c)\) yields exactly the same \(T_h^{\max}\) values, and hence the same Mpemba windows, as its classically correlated counterpart \(\hat{\rho}_{C}(m,T_c)\otimes \hat{\rho}_{P}(n-m,T_c)\) discussed in the previous subsection and summarized in Fig.~\ref{fig:multiC}. This already shows that multipartite entanglement, by itself, does not necessarily enhance the effect: for thermo-majorization under thermal operations, the relevant question is not the mere presence of entanglement, but whether the associated coherences can influence the zero-mode sector that controls population ordering.

\begin{figure} [t]
    \centering
    \includegraphics[width=.7\textwidth]{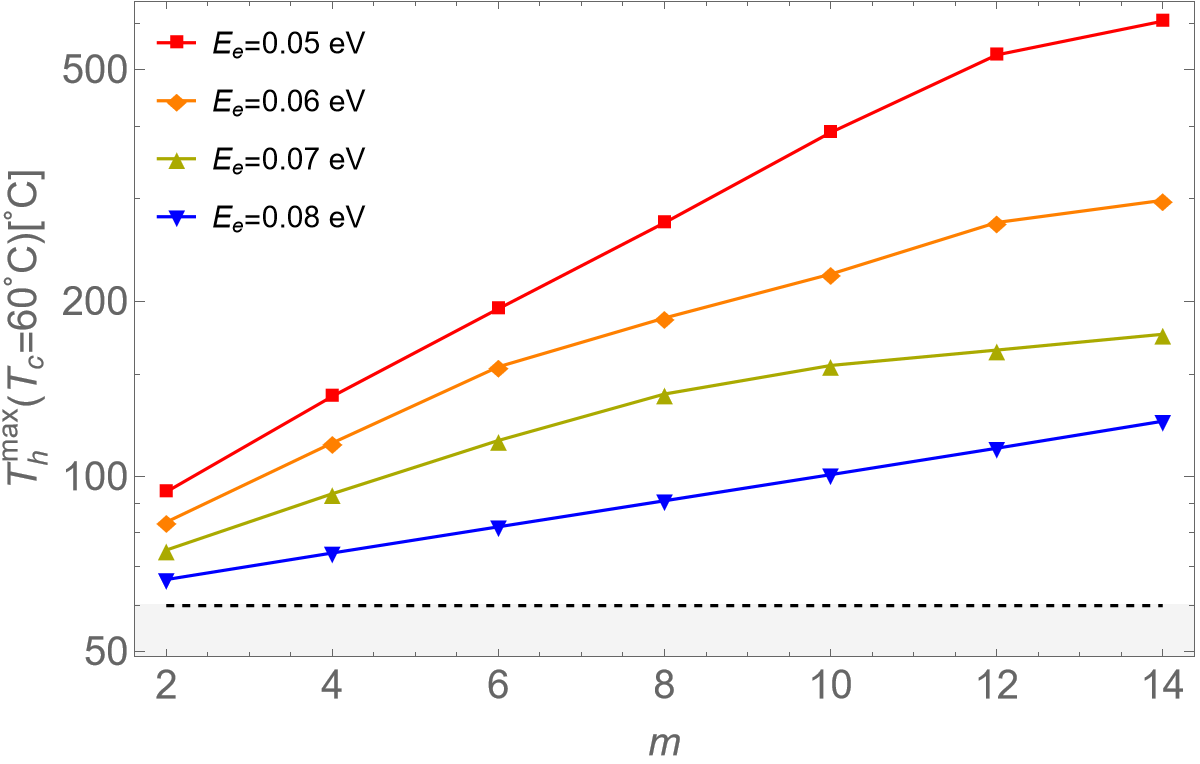}
    \caption{\justifying Maximal hot temperatures \(T_h^{\max}(T_c=60^\circ\mathrm{C})\) for partially dimerized entangled cold samples of the form \(\hat{\rho}_{\rm cold}=\hat{\rho}_{E}(2,T_c)^{\otimes m/2}\otimes \hat{\rho}_{P}(n-m,T_c)\) with fixed total size \(n=14\), where \(m\) qubits are arranged into \(m/2\) entangled pairs and the remaining \(n-m\) qubits are uncorrelated. Results are shown for different excitation gaps \(E_e-E_g\), using the maximal allowed coherence amplitude \(|\mu|=\sqrt{P_g(T_c)P_e(T_c)}\) in each entangled pair. The vertical axis is logarithmic. Over the plotted range, the curves appear approximately linear on this semi-log scale, indicating a rapid, roughly exponential growth of the Mpemba window as pairwise entanglement is distributed across more qubits.}
    \label{fig:SM:EntPairwisePartial}
\end{figure}

A more structured entangled family is obtained by distributing entanglement pairwise across the system,
\begin{equation}
  \hat{\rho}_{\rm cold}
=
\hat{\rho}_{E}(2,T_c)^{\otimes m/2}\otimes \hat{\rho}_{P}(n-m,T_c),
\end{equation}
so that \(m\) qubits are arranged into \(m/2\) entangled pairs while the remaining \(n-m\) qubits remain in the product thermal state.  
Unlike the GHZ-like multipartite family above, whose added coherences remain confined to nonzero modes of coherence and therefore do not affect the thermo-majorization ordering, the present pairwise-entangled family is consistent with the emergence of zero-mode coherence contributions in the composite system.  
Figure~\ref{fig:SM:EntPairwisePartial} shows the resulting \(T_h^{\max}(T_c=60^\circ\mathrm{C})\) for \(n=14\) as a function of \(m\).  
For all excitation gaps shown, \(T_h^{\max}\) increases monotonically with \(m\) over the plotted range.  
On the logarithmic vertical scale used in Fig.~\ref{fig:SM:EntPairwisePartial}, the curves appear approximately linear, consistent with a rapid, roughly exponential growth of the Mpemba window with the number of entangled pairs.  
Thus, when entanglement is distributed modularly rather than concentrated in a single collective block, its thermodynamic impact can accumulate across subsystems and substantially enlarge the accessible hot-temperature window.

\begin{figure} [h!]
    \centering
    \includegraphics[width=.7\textwidth]{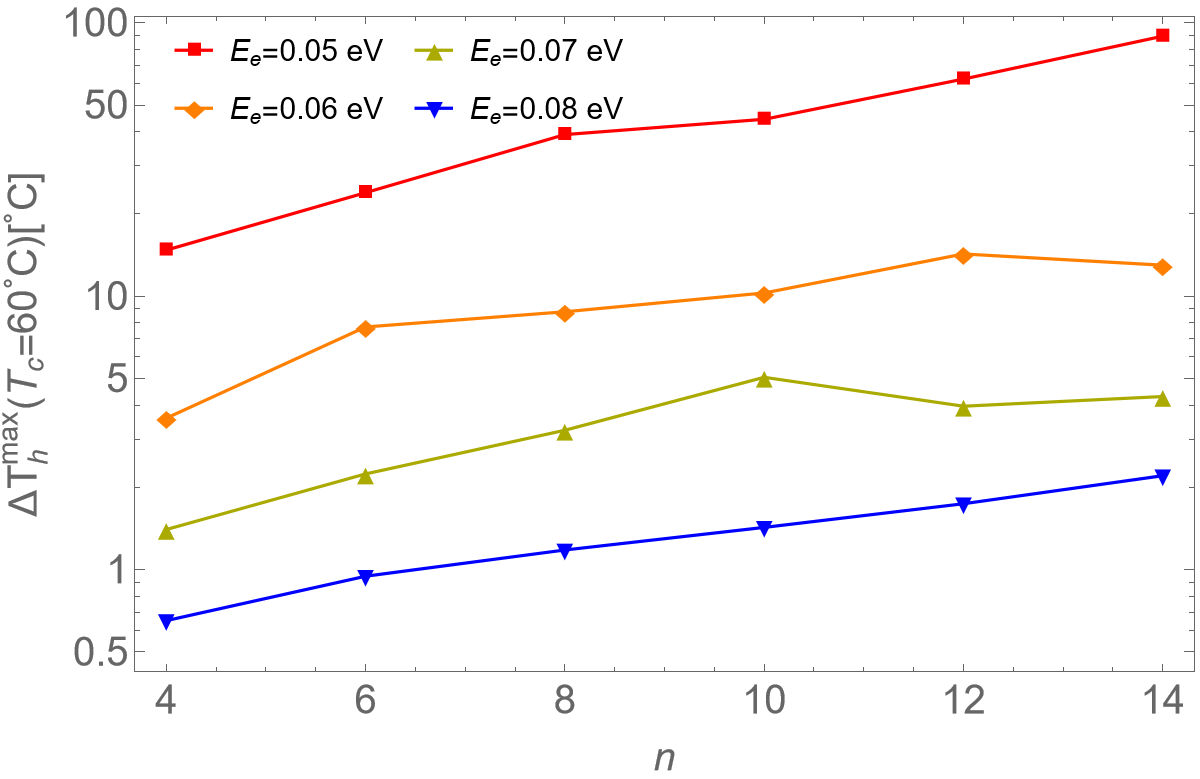}
    \caption{\justifying Quantum enhancement of the Mpemba window in fully dimerized systems, quantified by \(\Delta T_h^{\max}=T_{h,\mathrm{E}}^{\max}-T_{h,\mathrm{C}}^{\max}\), as a function of the even system size \(n\). Here \(\hat{\rho}_{\rm cold}=\hat{\rho}_{E}(2,T_c)^{\otimes n/2}\) is compared with its classically correlated counterpart \(\hat{\rho}_{C}(2,T_c)^{\otimes n/2}\), previously shown in Fig.~\ref{fig:SM:DimerizedMultiC}, at \(T_c=60^\circ\mathrm{C}\). Results are shown for different excitation gaps \(E_e-E_g\), using the maximal allowed coherence amplitude \(|\mu|=\sqrt{P_g(T_c)P_e(T_c)}\) in each entangled pair. The vertical axis is logarithmic. The enhancement vanishes at \(n=2\), becomes positive for \(n\ge 4\), and generally grows with system size, with the strongest advantage occurring for smaller excitation gaps.}
    \label{fig:SM:EntVsClassicalDimerized}
\end{figure}

The fully dimerized limit of this family,
\begin{equation}
  \hat{\rho}_{\rm cold}
=
\hat{\rho}_{E}(2,T_c)^{\otimes n/2},
\end{equation}
is the direct entangled counterpart of the classically dimerized configuration \(\hat{\rho}_{C}(2,T_c)^{\otimes n/2}\), whose behavior was shown in Fig.~\ref{fig:SM:DimerizedMultiC}.  
This case is particularly revealing because it allows a clean comparison between pairwise classical and pairwise quantum correlations at fixed local thermality.  
For \(n=2\), the two cases give exactly the same maximal hot temperature \(T_h^{\max}\), indicating that a single entangled pair does not outperform its classically correlated analogue in this setting.  
For larger even \(n\), however, the entangled dimerized family systematically exceeds the classically dimerized one.  
The quantum enhancement,
\begin{equation}
  \Delta T_h^{\max}
=
T_{h,\mathrm{E}}^{\max}
-
T_{h,\mathrm{C}}^{\max}
\end{equation}
is shown in Fig.~\ref{fig:SM:EntVsClassicalDimerized}.  
It vanishes at \(n=2\), becomes positive for \(n\ge 4\), and generally increases with system size, with the strongest enhancement occurring for smaller excitation gaps.  
For \(E_e=0.05~\mathrm{eV}\), the difference grows markedly with \(n\), whereas for \(E_e=0.08~\mathrm{eV}\) it remains modest but still clearly positive over the entire range considered.

This behavior is consistent with the emergence of additional zero-mode coherence contributions once multiple entangled pairs are combined into a larger composite system.  
While the coherence within a single pair does not by itself enlarge the Mpemba window beyond the classical dimerized benchmark, tensor-product replication of entangled pairs generates thermodynamically relevant structure that is absent in the classically correlated counterpart.  
In this sense, the advantage of the fully dimerized entangled family is not simply a pairwise effect repeated many times, but a genuinely many-body consequence of how quantum correlations are embedded across the composite Hilbert space.  
The contrast with the GHZ-like multipartite family above is especially instructive: collective multipartite entanglement confined to nonzero modes of coherence leaves the thermo-majorization criterion unchanged, whereas modularly distributed pairwise entanglement can generate a genuine quantum enhancement.

Taken together, these entangled-state examples reinforce the lesson already suggested by the classical analysis: the Mpemba effect in composite systems is governed not only by the amount of correlation but by its spectral and architectural organization.  
Entanglement can be thermodynamically inert when its coherence resides entirely outside the zero-mode sector, yet it can become operationally advantageous when distributed in a way that generates additional thermo-majorization-relevant structure in the composite system.  
This naturally motivates going beyond entanglement and examining quantum-correlated local thermal states whose relevant coherences need not be tied to entanglement itself.

\subparagraph{Discordant local thermal states.}
\begin{figure}[t]
    \centering
    \includegraphics[width=0.7\textwidth]{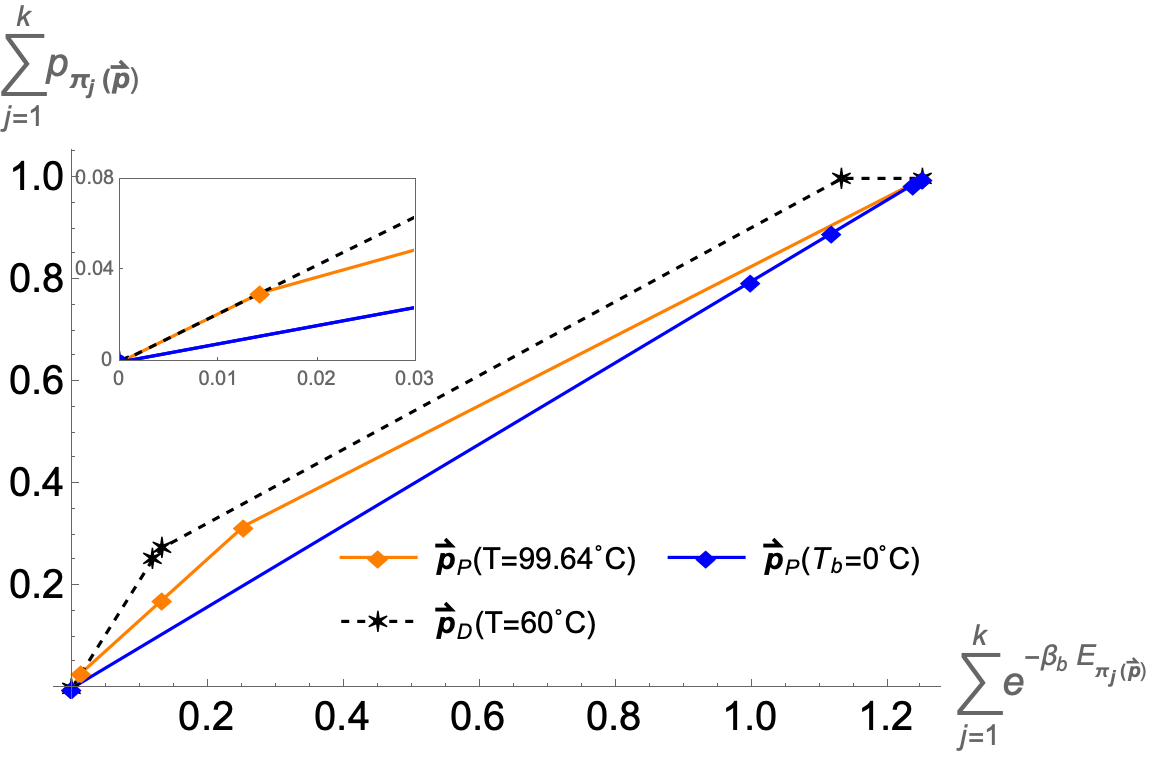}
    \caption{\justifying
    Thermo-majorization curves comparing deviations from equilibrium for locally thermal two-qubit states relative to the equilibrium product state at \(0^\circ\mathrm{C}\), \(\vec{p}_{P}(T_b=0^\circ\mathrm{C})\). For the discordant cold sample \(\vec{p}_{D}(T_c=60^\circ\mathrm{C})\), the maximal hot product state that remains thermo-majorized is \(\vec{p}_{P}(T_h^{\max}=99.62^\circ\mathrm{C})\). Thus, in the presence of quantum discord, a two-qubit sample at \(60^\circ\mathrm{C}\) can be farther from equilibrium than an uncorrelated sample at temperatures up to \(99.62^\circ\mathrm{C}\) during relaxation toward \(0^\circ\mathrm{C}\). The single-qubit energy levels are \(E_g=0\) and \(E_e=0.05~\mathrm{eV}\).
    }
    \label{fig:2qubitquant}
\end{figure}

A complementary scenario arises when locally thermal qubits share quantum correlations in separable joint states. In particular, we consider cold samples of the form
\begin{equation}
  \hat{\rho}_{\rm cold}
=
\hat{\rho}_{D}(m,T_c)\otimes \hat{\rho}_{P}(n-m,T_c),
\end{equation}
where \(m\) out of \(n\) qubits form a multipartite discordant state through
\begin{equation}
  \hat{\rho}_{D}(m,T_c)
=
\hat{\rho}_{P}(m,T_c)
+
\lambda (|ge\rangle\langle eg|)^{\otimes m/2}
+
\lambda^*(|eg\rangle\langle ge|)^{\otimes m/2},
\qquad
|\lambda|\le [P_g(T_c)P_e(T_c)]^{m/2}.
\end{equation}

Unlike the GHZ-like multipartite entangled family discussed above, whose added coherences remain confined to nonzero modes of coherence, all coherences in \(\hat{\rho}_{D}(m,T_c)\) lie in the zero mode. They can therefore dynamically interconvert with energy populations under thermal operations and directly influence the thermo-majorization ordering. For the discordant family considered here, however, the mere presence of zero-mode coherence is not by itself sufficient to generate a nontrivial Mpemba window. A necessary condition is that the zero-mode contribution changes the \(\beta\)-ordering relative to the locally thermal product state at the same temperature \(T_c\). If the \(\beta\)-ordering remains unchanged, the discordant thermo-majorization curve cannot fully overtake the equal-temperature product curve, and therefore cannot overtake any hotter product state with \(T_h>T_c\). A nontrivial discord-enabled Mpemba window thus requires not merely zero-mode coherence, but a genuine reordering of the \(\beta\)-ordered populations. 

The two-qubit case provides the clearest illustration of this mechanism. For \(\hat{\rho}_{\rm cold}=\hat{\rho}_{D}(2,T_c)\), the ordering reversal between the cold discordant sample and the hot product sample occurs when \((\beta_b-\beta_c)(E_e-E_g)\le \ln 2\) at maximal \(|\lambda|\). Setting \(E_g=0\) and \(E_e=0.05~\mathrm{eV}\), a discordant qubit pair prepared at \(T_c=60^\circ\mathrm{C}\) remains farther from equilibrium than a hot product pair up to \(T_h^{\max}=99.62^\circ\mathrm{C}\) during relaxation toward the bath temperature \(T_b=0^\circ\mathrm{C}\), as shown in Fig.~\ref{fig:2qubitquant}.

\begin{figure}[t]
    \centering
    \includegraphics[width=0.7\textwidth]{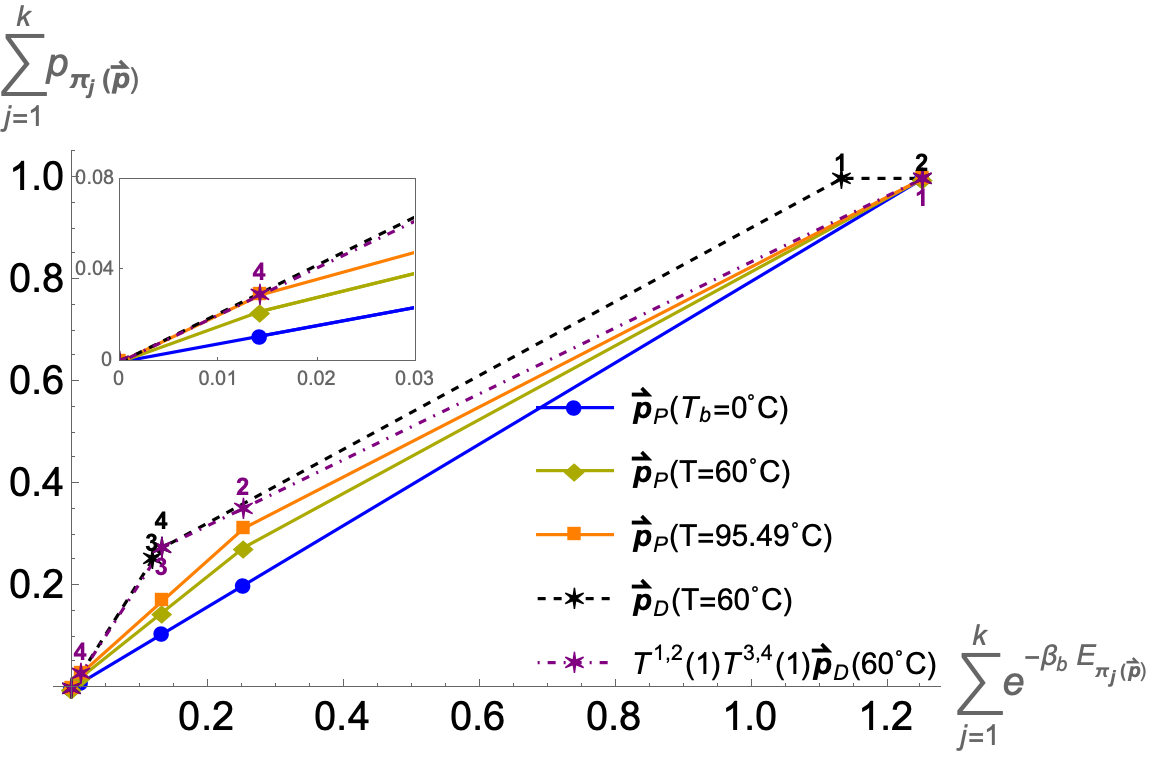}
    \caption{\justifying
    Thermo-majorization curves for the discordant two-qubit state \(\vec{p}_{D}(T_c=60^\circ\mathrm{C})\), its image under the full thermalization map \(T^{1,2}(1) \, T^{3,4}(1)\vec{p}_{D}\), and the maximal hot product state \(\vec{p}_{P}(T_h^{\max}=95.49^\circ\mathrm{C})\) thermo-majorized by the transformed state under Markovian thermal operations. The initial ordering of the discordant state is \((3,4,1,2)\), while after the action of \(T^{1,2}(1)\,T^{3,4}(1)\) it aligns with the product-state ordering \((4,3,2,1)\). The maximal hot product state touches the transformed discordant state at its first breakpoint (point 4). The single-qubit energy levels are \(E_g=0\) and \(E_e=0.05~\mathrm{eV}\).
    }
    \label{fig:conttmaj_pd}
\end{figure}

The operational class also matters. For the same discordant state \(\hat{\rho}_{\rm cold}=\hat{\rho}_{D}(2,60^\circ\mathrm{C})\), restricting the relaxation to Markovian thermal operations reduces the maximal hot temperature from \(99.62^\circ\mathrm{C}\) to \(95.49^\circ\mathrm{C}\). Figure~\ref{fig:conttmaj_pd} illustrates this reduction by comparing the initial discordant state, its image under the full thermalization map \(T^{1,2}(1) \, T^{3,4}(1)\vec{p}_{D}\), and the maximal product state thermo-majorized by the transformed distribution. Thus, while zero-mode coherence can enlarge the Mpemba window, the magnitude of the effect remains sensitive to the admissible dynamical structure.

The multiqubit behavior of this discordant family is qualitatively different from what one might naively expect from simply increasing the number of correlated qubits. Although zero-mode coherence directly contributes to thermo-majorization, the resulting Mpemba window does not grow with system size. Instead, the two-qubit case yields the widest window, while enlarging the discordant block progressively suppresses the effect.

\begin{figure}[t]
    \centering
    \includegraphics[width=0.7\textwidth]{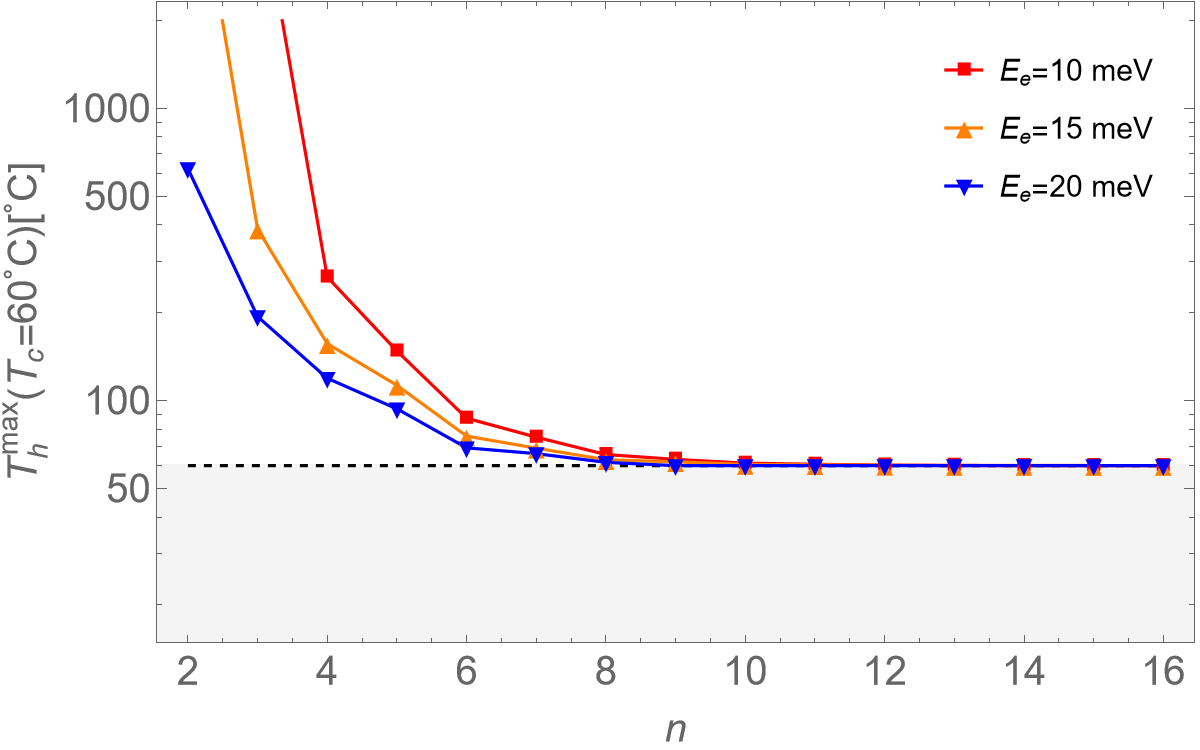}
    \caption{\justifying
    Maximum hot temperature \(T_h^{\max}\) for which the cold discordant state \(\hat{\rho}_{\rm cold} = \hat{\rho}_{D}(n,60^\circ\mathrm{C})\) thermo-majorizes the hot product state \(\hat{\rho}_{\rm hot} = \hat{\rho}_{P}(n,T_h)\), shown as a function of the total number of qubits \(n\) for several excitation gaps \(E_e-E_g\) (with \(E_g=0\)). For all energy gaps shown, the two-qubit case exhibits the widest Mpemba window, while increasing \(n\) progressively suppresses the effect and drives \(T_h^{\max}\) toward the trivial baseline \(T_c=60^\circ\mathrm{C}\) (dashed line). Thus, for this multipartite discordant family, zero-mode coherence is operationally active but its thermo-majorization advantage is not extensive under enlargement of the discordant sector.
    }
    \label{fig:SM:DiscordAll}
\end{figure}

\begin{figure}[t]
    \centering
    \includegraphics[width=0.7\textwidth]{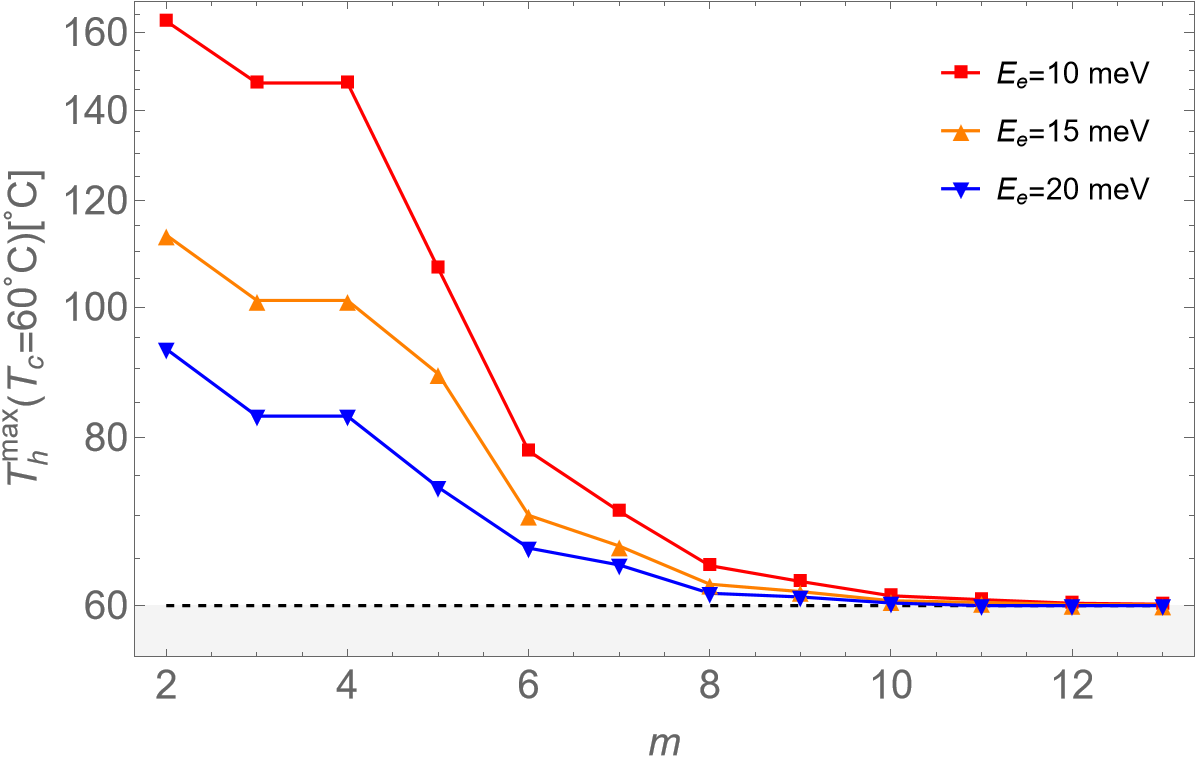}
    \caption{\justifying
    Maximum hot temperature \(T_h^{\max}\) for which the cold discordant state \(\hat{\rho}_{\rm cold} = \hat{\rho}_{D}(m,60^\circ\mathrm{C}) \otimes \hat{\rho}_{P}(13 - m,60^\circ\mathrm{C})\) thermo-majorizes the hot product state \(\hat{\rho}_{\rm hot} = \hat{\rho}_{P}(13,T_h)\), shown as a function of the discordant block size \(m\) for several excitation gaps \(E_e-E_g\) (with \(E_g=0\)). Even when the total system size is fixed, enlarging the discordant subset progressively narrows the Mpemba window and drives \(T_h^{\max}\) toward the trivial baseline \(T_c=60^\circ\mathrm{C}\) (dashed line).
    }
    \label{fig:SM:DiscordEmbedded}
\end{figure}

This suppression is observed both when the discordant structure spans the entire system (\(m=n\), with \(n\) increasing) and when the total system size is fixed but the discordant subset is enlarged within a locally thermal product background (\(m<n\), with \(m\) increasing). In both cases, the maximal hot temperature decreases toward the trivial baseline \(T_h^{\max}\approx T_c\), as shown in Figs.~\ref{fig:SM:DiscordAll} and \ref{fig:SM:DiscordEmbedded}. Thus, for this multipartite discordant family, zero-mode coherence is operationally active but its thermo-majorization advantage is not extensive under enlargement of the discordant sector.

The same large-system limit is obtained when only a single pair remains discordant while all additional qubits stay uncorrelated, i.e.,
\begin{equation}
  \hat{\rho}_{\rm cold}
=
\hat{\rho}_{D}(2,T_c)\otimes \hat{\rho}_{P}(n-2,T_c).
\end{equation}
In this case, increasing the total number of spectator qubits also suppresses the effect: the maximal hot temperature again decreases toward the trivial baseline \(T_h^{\max}\approx T_c\). Thus, even though the operationally active zero-mode coherence is localized in a single discordant pair, its thermo-majorization advantage is progressively diluted when embedded into a sufficiently large locally thermal product background.

A sharper contrast is obtained for a dimerized discord pattern,
\begin{equation}
  \hat{\rho}_{\rm cold}
=
\hat{\rho}_{D}(2,T_c)^{\otimes n/2},
\end{equation}
in which the system is partitioned into \(n/2\) independent discordant pairs. In this case, the maximal hot temperature is exactly equal to that of the two-qubit discordant state for all even \(n\). For instance, with \(E_g=0\) and \(E_e=\{0.05,0.06,0.07,0.08\}\,\mathrm{eV}\), one obtains the \(n\)-independent values
\[
T_h^{\max}=\{99.63^\circ\mathrm{C},\,82.55^\circ\mathrm{C},\,72.17^\circ\mathrm{C},\,65.23^\circ\mathrm{C}\},
\]
respectively. Thus, unlike the multipartite discordant family and the single-pair embedded family, a replicated pairwise discord structure is exactly size-stable: the critical temperature is fixed entirely by the two-qubit discordant building block and is neither amplified nor diluted by tensor-product replication.

These examples reinforce a central lesson of the correlation-enabled Mpemba effect: the presence of quantum correlations alone is not decisive; their operational impact depends crucially on how they are distributed across subsystems and across coherence modes, with zero-mode participation playing the central role in the discordant cases considered here.

\section{Mpemba Window in Single-Molecule Water Model}
\label{subsec:SM:water}

\subsection{A single-molecule effective Hamiltonian}
\label{sec::SI_water_model}

Our resource-theoretic analysis of water addresses an operational aspect of the Mpemba problem that is not naturally resolved by ensemble-based statistical analyses of bulk water or by phenomenological master equations developed for cooling dynamics. To this end, we construct an effective Hamiltonian whose Gibbs state provides a temperature-dependent local reference for the distance-to-equilibrium partial ordering at the level of single molecules. This choice is deliberate: within the thermal-operation framework presented in the main text, the relevant comparisons are between multipartite initial states built from local Gibbs marginals. A temperature-dependent single-molecule Hamiltonian is therefore the minimal physically interpretable input needed to construct correlated and uncorrelated multi-molecule states without committing to a separate many-body equilibrium theory of water.

Accordingly, the model should not be viewed as a competitor to established microscopic or many-body water models. It is instead a reduced local surrogate tailored to a narrower operational question: whether classical correlations between locally thermal water motifs can enable a thermomajorization-based Mpemba window.

To this end, we represent a single water molecule by seven effective local configurations,
\begin{equation}
  \{0_{\rm V},0_{\rm L},1_{\rm L},2_{\rm L},3_{\rm L},4_{\rm L},4_{\rm I}\},
\end{equation}
where the label indicates the number of hydrogen bonds (HBs), and the subscript distinguishes physically different local motifs at the two extremes of the bonding spectrum. In particular, \(0_{\rm V}\) denotes a vapor-like unbonded configuration with large effective entropic weight, while \(0_{\rm L}\) denotes an isolated but still liquid-like local environment. Similarly, \(4_{\rm L}\) denotes a fully bonded but distorted liquid-like local configuration, whereas \(4_{\rm I}\) represents an idealized ice-like tetrahedral environment.

The seven-level structure is the minimal extension of the natural HB-count basis \(N=0,\dots,4\). We do not proliferate levels indiscriminately; rather, we resolve only the two endpoints where a single HB count is known to conflate physically distinct local environments. The split states \(4_{\rm I}/4_{\rm L}\) and \(0_{\rm V}/0_{\rm L}\) are therefore not introduced as literal phase fractions of bulk water, but as effective local motifs that allow the single-molecule Gibbs state to encode the dominant restructuring of hydrogen-bond environments near freezing-like and vaporization-like regimes.

For a configuration with \(N\in\{0,1,2,3,4\}\) HBs in the liquid-like sector, we assign the effective energy
\begin{equation}
  E_N(T)= N\epsilon_0(1-\alpha T)+\frac{\gamma}{2}N(N+1),
\end{equation}
where \(\epsilon_0\) sets the baseline HB energy scale, \(\alpha\) captures thermal weakening of HBs, and \(\gamma\) encodes effective many-bond crowding or cooperativity. We then refine the two extremes of the bonding manifold. At the fully bonded end, the ice-like tetrahedral configuration is stabilized relative to the distorted liquid-like one according to
\begin{equation}
    E_{4_{\rm I}}(T)=E_{4_{\rm L}}(T)-\frac{\Delta q}{1+e^{(T_{L\leftrightarrow I}-T)}}.
\end{equation}
At the unbonded end, the vapor-like configuration is assigned an enhanced effective weight
\begin{equation}
    w_{\rm vap}(T)=1+\frac{\Delta_{\rm vap}}{1+e^{(T-T_{L\leftrightarrow V})}},
\end{equation}
which phenomenologically captures the much larger effective multiplicity of vapor-like local configurations. The resulting effective Hamiltonian is
\begin{equation}
    \hat{H}_{\rm H_2O}(T)=\sum_{j} E_j(T)\,|E_j\rangle\langle E_j|,
\end{equation}
with \(j\in\{0_{\rm V},0_{\rm L},1_{\rm L},2_{\rm L},3_{\rm L},4_{\rm L},4_{\rm I}\}\), and with the understanding that the \(0_{\rm V}\) sector carries the additional weight \(w_{\rm vap}(T)\) in the partition function. The corresponding single-molecule Gibbs state is then defined by the normalized effective Gibbs weights \(P_j(T)\) of these seven local configurations.

Throughout, the purpose of this construction is not to reproduce the full equilibrium thermodynamics of water, but to provide a compact temperature-dependent local Gibbs model that captures coarse-grained restructuring of HB environments in a form compatible with thermal operations and thermomajorization.

\subsection{Parameterization, local structural diagnostics, and physical motivation}
\label{sec::SI_water_fit}
\begin{figure}[t]
    \centering
    \includegraphics[width=0.7\textwidth]{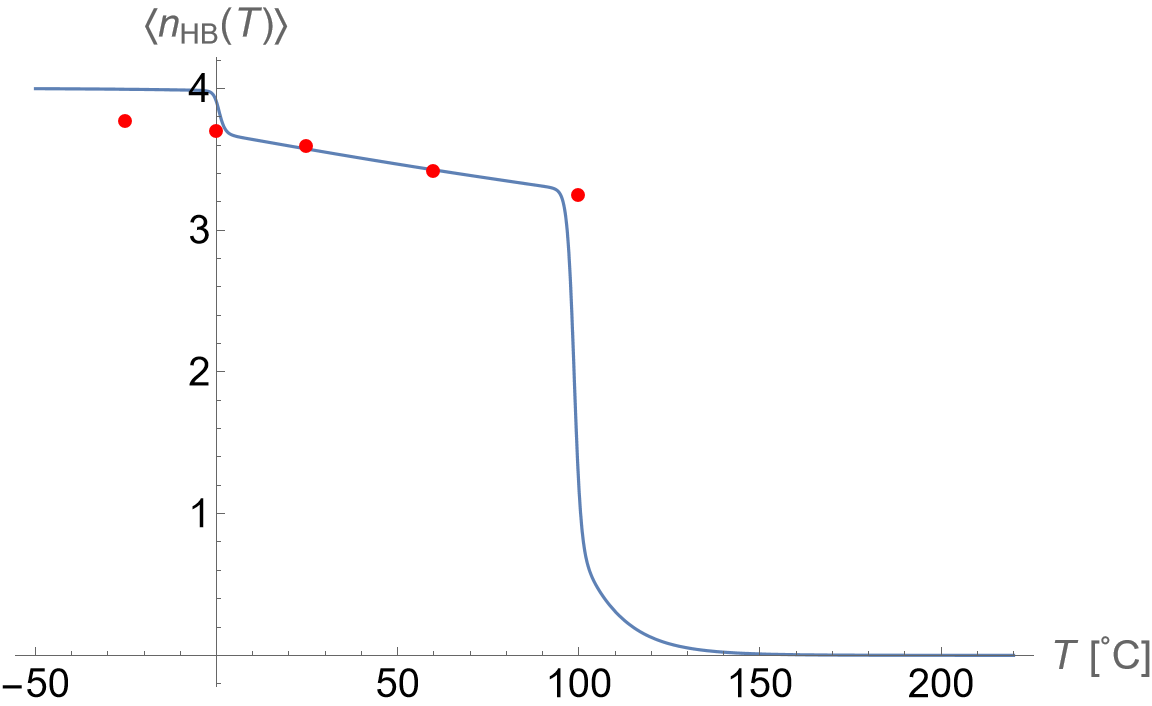}
    \caption{\justifying Temperature dependence of the mean hydrogen-bond number \(\langle n_{\rm HB}(T)\rangle\) in the seven-level single-molecule effective model (solid curve). Red points denote the TIP4P-based reference values used to calibrate the liquid-regime trend. The model is designed to reproduce the smooth decrease of \(\langle n_{\rm HB}(T)\rangle\) across the liquid window while also generating effective low- and high-temperature crossovers associated with ice-like and vapor-like local restructuring. These crossovers should be interpreted as features of the surrogate local Gibbs model, not as literal bulk phase boundaries.
    }
    \label{fig::SI_HBfit}
\end{figure}

The five model parameters were chosen so that the single-molecule Gibbs state reproduces key temperature-dependent \emph{local} structural trends associated with water:
\[
\epsilon_0=-0.66931~{\rm eV},\qquad
\alpha=7.097\times10^{-4}~{\rm K}^{-1},\qquad
\gamma=0.13436~{\rm eV},
\]
\[
\Delta q=0.09519~{\rm eV},\qquad
\Delta_{\rm vap}=6.03\times10^9.
\]
These parameters are not introduced as a unique microscopic derivation from TIP4P or any other detailed water model. Rather, TIP4P data reported in Ref.~\cite{jorgensen1985watersim} serve as a calibration benchmark: the parameters were chosen so that the resulting single-molecule Gibbs state reproduces the temperature dependence of the mean HB number \(\langle n_{\rm HB}(T)\rangle\) in the liquid regime, while also enforcing physically sensible low- and high-temperature restructuring near freezing-like and vaporization-like conditions. The same parameter set also reproduces a reasonable temperature dependence of a coarse-grained tetrahedrality measure \(\langle q(T)\rangle\).

\begin{figure}[t]
    \centering
    \includegraphics[width=0.7\textwidth]{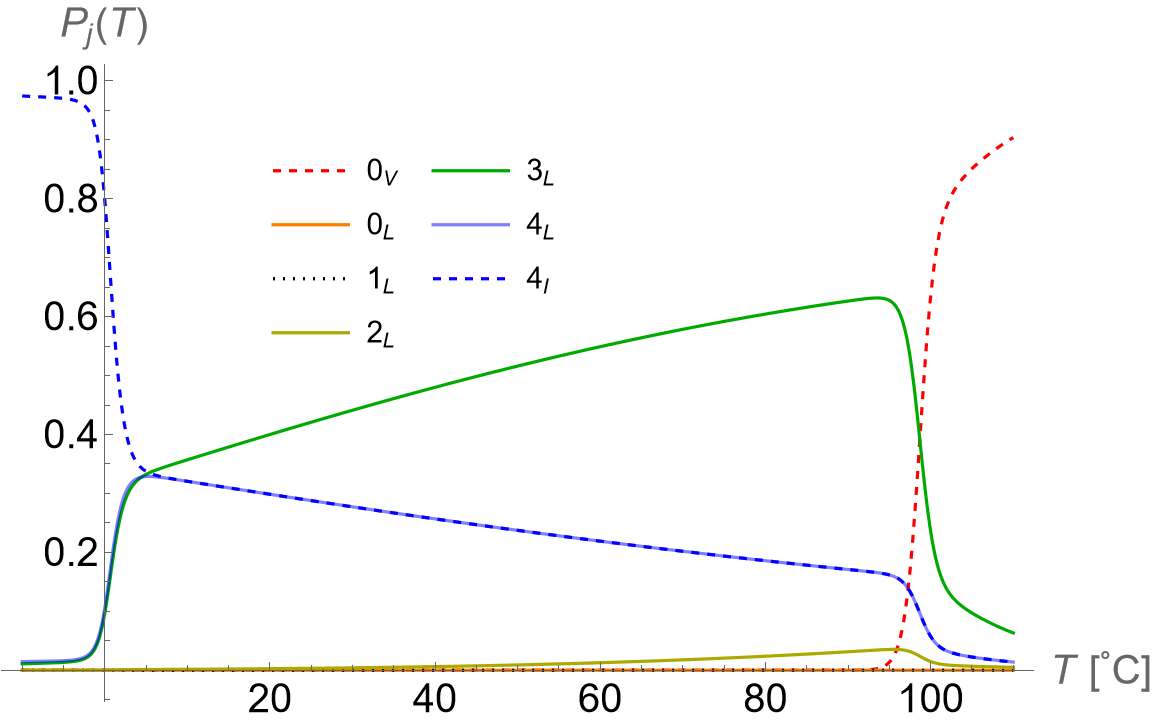}
    \caption{\justifying
    Effective Gibbs weights \(P_j(T)\) of the seven local configurations in the single-molecule water model. At low temperatures, the ice-like four-bond sector \(4_{\rm I}\) dominates and gradually transfers weight to the distorted liquid-like sectors \(4_{\rm L}\) and \(3_{\rm L}\) as temperature increases. At high temperatures, the vapor-like unbonded sector \(0_{\rm V}\) becomes dominant. These curves should be interpreted as probabilities of effective local configurations in the surrogate Gibbs model, not as literal bulk phase fractions.
    }
    \label{fig::SI_populations}
\end{figure}

Figure~\ref{fig::SI_HBfit} shows that the resulting model reproduces the monotonic decrease of \(\langle n_{\rm HB}(T)\rangle\) across the liquid regime, while generating smooth low- and high-temperature crossovers rather than sharp discontinuities. Figure~\ref{fig::SI_populations} shows the corresponding effective Gibbs weights \(P_j(T)\). At low temperatures, the \(4_{\rm I}\) sector is strongly favored, but smoothly transfers weight to \(4_{\rm L}\) and \(3_{\rm L}\) as the temperature rises. Across the liquid-like window, the dominant competition is between \(4_{\rm L}\) and \(3_{\rm L}\), with smaller contributions from lower-bond liquid-like motifs. At high temperatures, the \(0_{\rm V}\) sector becomes dominant, providing a smooth surrogate of vaporization-like local restructuring. These quantities should be interpreted as Gibbs weights of effective \emph{local configurations}, not as literal equilibrium phase fractions.

This distinction is important for the main operational claim. The present model is not used to infer actual nonequilibrium cooling trajectories of water. Rather, it defines a physically motivated family of local thermal reference states against which one can ask whether correlated and uncorrelated multipartite preparations are ordered differently under thermomajorization.

The correlated preparations used in the main text are also physically motivated rather than purely abstract ans\"atze. In hydrogen-bond networks, transient proton delocalization can generate short-lived quantum correlations between neighboring water molecules or local motifs, as shown in earlier microscopic open-system models~\cite{Pusuluk_2018_PRSA,Pusuluk_2019_PRSA}. Rapid molecular motion is then expected to suppress such coherences on longer timescales, leaving behind effectively classical correlations in coarse-grained local descriptors such as hydrogen-bond environments. From this perspective, the classically correlated states used below should be viewed as phenomenological coarse-grained surrogates of such locally structured preparations.

\subsection{Operational liquid-like window and bath-temperature choice}
\label{sec::SI_water_bath}
\begin{figure}[t]
    \centering
    \includegraphics[width=0.7\textwidth]{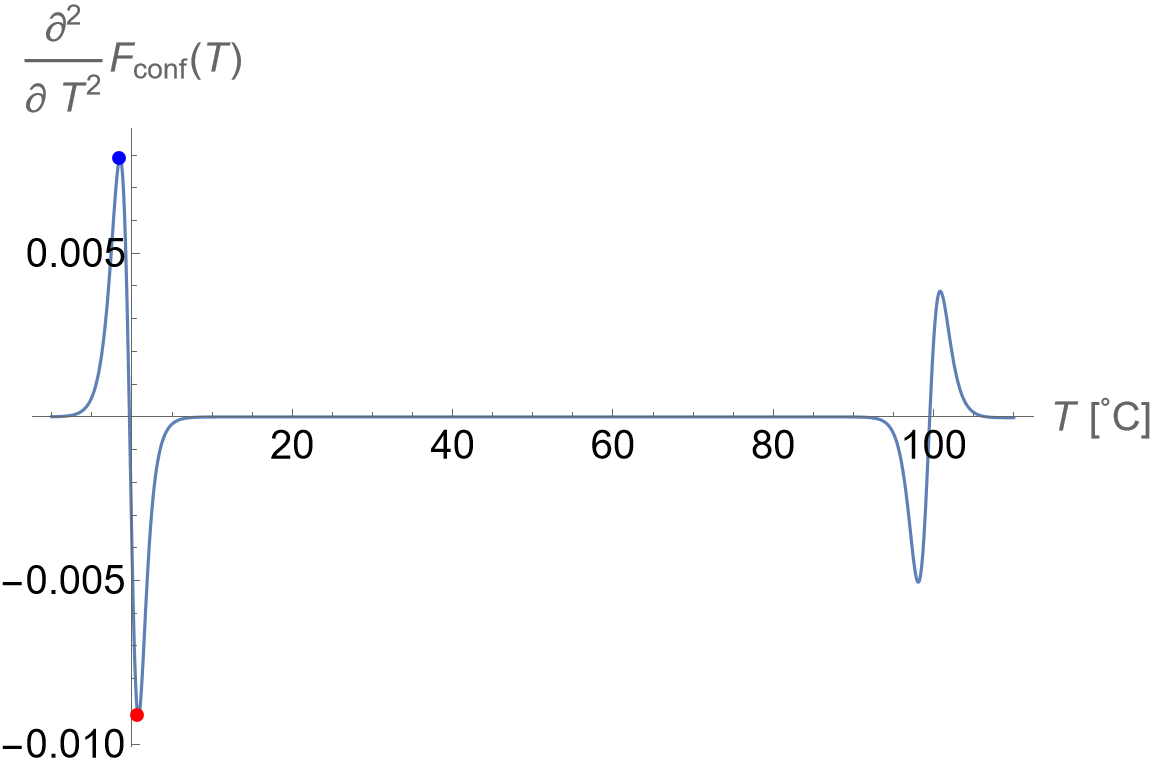}
    \caption{\justifying
    Second temperature derivative of the configurational free energy,
    \(\partial_T^2 F_{\rm conf}(T)\), in the seven-level single-molecule effective model. The low-temperature anomaly is used as an operational marker for the onset of the liquid-like sector within the effective description, motivating the bath-temperature choice \(T_b=0.8^\circ\mathrm{C}\) (red dot) adopted in the thermomajorization analysis. The extrema are not interpreted as literal predictions of bulk phase-transition temperatures, but as crossover diagnostics internal to the surrogate model.
    }
    \label{fig::SI_Fcurv}
\end{figure}

Because the thermomajorization criterion compares initial states relative to a bath Gibbs state, one must specify a bath temperature \(T_b\) for the water model. To avoid placing the bath inside the low-temperature crossover region between ice-like and liquid-like local configurations, we choose
\[
T_b=0.8^\circ{\rm C},
\]
which we take as the lowest temperature on the liquid-like side of the effective model.

Operationally, as shown in Fig.~\ref{fig::SI_Fcurv}, this choice is diagnosed from the low-temperature anomaly in the curvature of the configurational free energy,
\begin{equation}
  F_{\rm conf}(T)=-k_B T\ln Z(T),
\end{equation}
and, consistently, from the corresponding structural curvature diagnostics such as \(\partial_T^2\langle n_{\rm HB}(T)\rangle\). The associated extrema do not represent literal predictions of a bulk phase-transition point. Rather, they serve as operational markers for the onset and eventual breakdown of the liquid-like sector within the effective single-molecule description. Choosing \(T_b=0.8^\circ{\rm C}\) therefore ensures that the Gibbs reference state entering the thermomajorization test is evaluated within the same liquid-like sector used to parameterize the local HB statistics.

This operational reading should be kept in mind throughout: the resulting water Mpemba windows are windows of \emph{thermomajorization ordering} within the effective model, not direct predictions of experimentally observed cooling-rate inversions.

\subsection{Construction of correlated and uncorrelated multi-molecule states}
\label{sec::SI_water_states}

Given the single-molecule Gibbs state \(\hat{\rho}_{\rm th}(T,\hat{H}_{\rm H_2O})\), we define the \(n\)-molecule product state
\begin{equation}
  \hat{\rho}_{P}(({\rm H_2O})_n,T)
=
\hat{\rho}_{\rm th}(T,\hat{H}_{\rm H_2O})^{\otimes n},
\end{equation}
and the fully classically correlated state
\begin{equation}
\hat{\rho}_{C}(({\rm H_2O})_n,T)
=
\sum_j P_j(T)\,\bigl(|E_j\rangle\langle E_j|\bigr)^{\otimes n},
\end{equation}
where \(P_j(T)\) denotes the normalized effective Gibbs weight of the \(j\)-th local configuration. By construction, \(\hat{\rho}_{C}\) and \(\hat{\rho}_{P}\) share the same single-molecule thermal marginals, but \(\hat{\rho}_{C}\) contains classical correlations in the local energy basis.

For the mixed correlated/product preparations studied in the main text, we define
\begin{equation}\label{eq:ColdCorrW}
\hat{\rho}_{\rm cold}
=
\hat{\rho}_{C}(({\rm H_2O})_m,T_c)\otimes
\hat{\rho}_{P}(({\rm H_2O})_{n-m},T_c),
\end{equation}
and compare them against the uncorrelated hotter preparation
\begin{equation}\label{eq:HotProW}
\hat{\rho}_{\rm hot}
=
\hat{\rho}_{P}(({\rm H_2O})_n,T_h).
\end{equation}
As in the qubit examples of the main text, the question is whether the colder correlated state can remain farther from equilibrium than the hotter product state, in the precise sense determined by thermomajorization with respect to the bath Gibbs state at temperature \(T_b\). Whenever this holds while \(T_h>T_c\), the pair \((T_c,T_h)\) lies inside the corresponding water Mpemba window.

As in the qubit setting, correlations are therefore treated as potentially \emph{enabling} resources, not as sufficient ones. Whether a window exists depends not only on the presence of correlations, but also on how the correlated weight is distributed across the effective local spectrum.

\subsection{Additional view of the water Mpemba windows}
\label{sec::SI_water_heatmap}
\begin{figure}[t]
    \centering
    \includegraphics[width=0.5\textwidth]{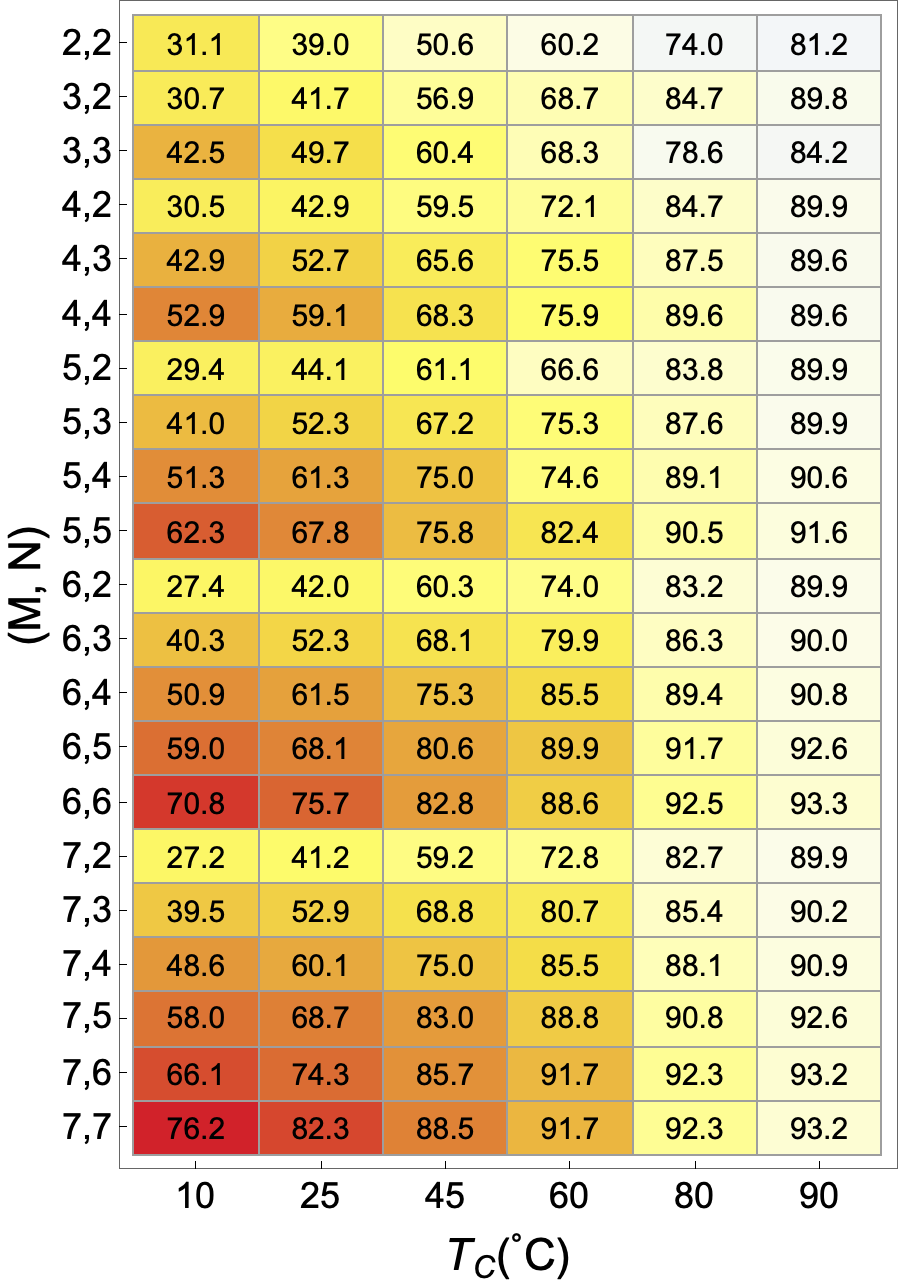}
    \caption{\justifying
    Heat-map representation of the maximal hot temperature \(T_h^{\max}(T_c)\) for which the colder correlated state \eqref{eq:ColdCorrW} thermo-majorizes the hotter product state \eqref{eq:HotProW} with respect to the bath Gibbs state at \(T_b=0.8^\circ\mathrm{C}\). This provides a visual complement to Table~\ref{tab:watercmn} in the main text. The plot makes clear that the water Mpemba window is strongly state-dependent: increasing the correlated fraction \(m\) can enlarge the admissible hot-temperature range, but the effect is not strictly monotonic and becomes constrained near the upper end of the liquid-like sector of the effective model.
    }
    \label{fig::SI_water_heatmap}
\end{figure}

Table~\ref{tab:watercmn} of the main text reports the \emph{width} of the water Mpemba window,
\begin{equation}
  \Delta T_{\rm M}=T_h^{\max}(T_c)-T_c,
\end{equation}
for representative values of \(T_c\), \(n\), and \(m\). The same data can be visualized more directly by plotting the maximal admissible hot temperature \(T_h^{\max}(T_c)\) itself as a heat map over the \((n,m)\) sector (see Fig.~\ref{fig::SI_water_heatmap}). This representation makes several trends particularly transparent.

First, for fixed \((n,m)\), the maximal hot temperature generally increases with \(T_c\), but the \emph{window width} \(\Delta T_{\rm M}\) simultaneously tends to shrink because the baseline temperature \(T_c\) is itself increasing. This explains why the main-text table shows the largest Mpemba windows at low \(T_c\), even though the corresponding \(T_h^{\max}\) values are not maximal there.

Second, for fixed \(n\), increasing the correlated fraction \(m\) often raises \(T_h^{\max}\), consistent with the intuition that stronger classical correlations can push the colder state farther from equilibrium. However, this dependence is not strictly monotonic in all sectors, reflecting the same broader lesson emphasized throughout the paper: correlations can be operationally relevant without being sufficient on their own, because the detailed distribution of correlated weight across the spectrum also matters.

Third, at high \(T_c\), many entries approach a common ceiling near the upper edge of the liquid-like window of the effective model. This saturation indicates that the disappearance of the Mpemba window at high temperatures is controlled not only by the weakening of the correlation advantage, but also by the finite temperature range over which liquid-like local Gibbs states remain meaningful in the present effective description.

Overall, the heat-map view reinforces the main conclusion of the water section: even within a deliberately minimal single-molecule description, classical correlations between local thermal water motifs can generate colder-but-farther configurations, but only in a strongly state-dependent and temperature-dependent manner. This sensitivity provides a natural operational rationale for why Mpemba-like signatures in water can be sporadic, protocol-dependent, and sometimes contradictory across different preparation schemes.

\end{document}